%% file: main.tex
\documentclass[acmsmall]{acmart}

\setcopyright{none} 
\renewcommand\footnotetextcopyrightpermission[1]{}

\usepackage{enumitem}

\usepackage{amsmath,amsfonts}
\usepackage{algorithmic}
\usepackage{graphicx}
\usepackage{xcolor}
\usepackage{url}
\usepackage{booktabs}
\usepackage{mdframed}
\usepackage{pifont}
\usepackage{multirow}
\usepackage{diagbox}
\usepackage{makecell}
\usepackage{array}
\usepackage{subcaption} 

\newcommand{\revise}[1]{{#1}}
\newcommand{\rerevise}[1]{{#1}}

\newcommand{\parabf}[1]{\medskip\noindent\textbf{#1}}

\newcommand{\paraf}[1]{\noindent\textbf{#1}}

\newcounter{finding}
\newcommand{\finding}[2]{\refstepcounter{finding}
\vspace{2.3mm}
 \begin{mdframed}[linecolor=gray,roundcorner=12pt,backgroundcolor=gray!15,linewidth=3pt,innerleftmargin=2pt, leftmargin=0cm,rightmargin=0cm,topline=false,bottomline=false,rightline = false]
  \textbf{Finding \arabic{finding}:} #1
 \end{mdframed}
 \label{#2}
 \vspace{2.3mm}
}

\usepackage{listings}
\usepackage{xcolor} 
\definecolor{codegray}{gray}{0.5} 
\AtBeginDocument{%
  }

\copyrightyear{2025}
\acmYear{2025}
\acmDOI{}

\begin{document}

\title{A First Look at Bugs in LLM Inference Engines}

\author{Mugeng Liu}
\affiliation{%
  \institution{School of Computer Science, Peking University}
  \city{Beijing}
  \country{China}}
\email{lmg@pku.edu.cn}

\author{Siqi Zhong}
\affiliation{%
  \institution{Institute for Artificial Intelligence, Peking University}
  \city{Beijing}
  \country{China}}
\email{siqizhong25@stu.pku.edu.cn}

\author{Weichen Bi}
\affiliation{%
  \institution{Institute for Artificial Intelligence, Peking University}
  \city{Beijing}
  \country{China}}
\email{biweichen@pku.edu.cn}

\author{Yixuan Zhang}
\affiliation{%
  \institution{School of Computer Science, Peking University}
  \city{Beijing}
  \country{China}}
\email{zhangyixuan.6290@pku.edu.cn}

\author{Zhiyang Chen}
\affiliation{%
  \institution{Institute for Artificial Intelligence, Peking University}
  \city{Beijing}
  \country{China}}
\email{zhiyangchen@stu.pku.edu.cn}

\author{Zhenpeng Chen}
\authornote{Corresponding authors.}
\affiliation{%
  \institution{School of Software, Tsinghua University}
  \city{Beijing}
  \country{China}}
\email{zpchen@tsinghua.edu.cn}

\author{Xuanzhe Liu}
\affiliation{%
  \institution{School of Computer Science, Peking University}
  \city{Beijing}
  \country{China}}
\email{xzl@pku.edu.cn}

\author{Yun Ma}
\authornotemark[1]
\affiliation{%
  \institution{Institute for Artificial Intelligence, Peking University}
  \city{Beijing}
  \country{China}}
\email{mayun@pku.edu.cn}

\renewcommand{\shortauthors}{Liu et al.}

\begin{abstract}
Large language model-specific inference engines (in short as \emph{LLM inference engines}) have become a fundamental component of modern AI infrastructure, enabling the deployment of LLM-powered applications (LLM apps) across cloud and local devices. 
Despite their critical role, LLM inference engines are prone to bugs due to the immense resource demands of LLMs and the complexities of cross-platform compatibility. 
However, a systematic understanding of these bugs remains lacking. 
To bridge this gap, we present the first empirical study on bugs in LLM inference engines.
We mine official repositories of 5 widely adopted LLM inference engines, constructing a comprehensive dataset of 929 real-world bugs. 
\revise{Through a rigorous open coding process, we analyze these bugs to uncover their symptoms, root causes, commonality, fix effort, fix strategies, and temporal evolution.}
Our findings reveal six bug symptom types and a taxonomy of 28 root causes, shedding light on the key challenges in bug detection and location within LLM inference engines. 
Based on these insights, we propose a series of actionable implications for researchers, inference engine vendors, and LLM app developers, along with general guidelines for developing LLM inference engines.
\end{abstract}



\maketitle

\input{sections/introduction}

\input{sections/background}

\input{sections/methodology}

\input{sections/symptom}

\input{sections/root_cause}

\input{sections/commonality}

\input{sections/repair_effort}

\input{sections/fix_strategies}

\input{sections/temporal}
\input{sections/discussion}
\input{sections/relatedwork}

\input{sections/conclusion}

\section*{Acknowledgment}

This work was supported by Beijing Natural Science Foundation (No. L253005), the National Natural Science Foundation of China (No. 62325201), the Key Laboratory of High Confidence Software Technologies (Peking University), the Ministry of
Education, and the Center for Data Space Technology and System, Peking University.


\bibliographystyle{ACM-Reference-Format}
\bibliography{main.bbl}

\end{document}

%% file: sections/introduction.tex
\section{Introduction}
\label{sec:introduction}

Large language model-specific inference engine (abbreviated as LLM inference engine) has emerged as a vital component of modern artificial intelligence infrastructure. 
\revise{An LLM inference engine is specialized software designed to run pre-trained LLMs, enabling their prediction or decision generation process~\cite{wiki_inference_engine}}.
These engines are specially optimized for LLM inference, offering LLM-specific optimizations tailored to reduce resource consumption and enhance inference speed. 
LLM inference engines have significantly boosted the adoption of LLM-powered applications (LLM apps). 
While LLM inference engines deliver fast, high-quality responses in cloud environments~\cite{chatgpt}, they also support local deployment on personal devices like Windows, Linux, and MacOS PCs for personal LLM apps such as Ollama~\cite{ollama}.

LLM inference engine is complexly constructed and highly bug-prone due to the immense resource demands of LLMs and the challenges of cross-platform compatibility.
First, due to the vast number of parameters in LLMs, LLM inference is both compute-intensive and memory-hungry.
\revise{For instance, running Llama-2-70B on a single GPU typically requires more than 80 GB of GPU memory, which makes it infeasible without model sharding or quantization techniques. Even smaller-scale engines such as Llama.cpp require more than 16 GB of RAM to execute models with 13B parameters efficiently.}
To enhance inference efficiency, LLM inference engines employ numerous complex optimizations to reduce the memory consumption and enhance the inference latency, such as paged attention, continuous batching, and speculative decoding.
\revise{For instance, vLLM employs paged attention, which organizes the Key-Value (KV) cache into non-contiguous pages.
The KV cache is a critical component for storing attention states during autoregressive generation of LLMs. 
This approach requires fine-grained memory management to reduce memory fragmentation caused by variable-length sequences, thereby reducing memory waste, which is a common bottleneck in LLM inference.}
Second, the engines should adapt to a wide range of application settings.
They need to support various LLM architectures across different environments, including multi-node clusters, single nodes with multi-GPUs, single GPU servers, PCs, mobile devices, IoT devices, and browsers.
The combination of the complex optimizations and diverse environment adoption potentially leads to more bugs, which can severely impact the availability, performance, and user experience of LLM apps.

Bugs in AI systems have been extensively studied. 
Previous work has empirically investigated bugs of deep learning (DL) systems across stages, including the learning stage~\cite{zhang2018empirical,humbatova2020taxonomy}, the compiling stage~\cite{shen2021comprehensive,guan2023comprehensive}, and the deployment stage~\cite{chen2020comprehensive,chen2021empirical}.
However, a systematic understanding and characterization of bugs in LLM inference engines remains lacking.
This knowledge is urgently needed as LLMs have become the dominant AI paradigm and more prone to bugs compared to traditional DL systems.
 
To bridge this gap, we present the first empirical study on bugs in LLM inference engines. 
Given the increasing prevalence of LLM apps, this research is of great significance. 
Specifically, we aim to address the following research questions (RQs):

\revise{
\parabf{(RQ1) Symptoms:} What are the symptoms of bugs and their distribution?
Understanding the symptoms helps clarify the impacts of bugs and guides the efficient setting of test oracles.

\paraf{(RQ2) Root Causes:} What are the root causes of bugs and their distribution?
The root causes help identify vulnerabilities and guide bug detection and localization.
}

\paraf{(RQ3) Commonality:} Do bugs follow a common pattern across engines?
Investigating commonality reveals cross-engine generalization of bug patterns and shared vulnerabilities to guide the testing methods.

\revise{
\paraf{(RQ4) Fix Effort:} Which bugs require more effort to fix?
Investigating fix effort quantifies the maintenance burden associated with different bugs, which could help prioritize testing and development efforts.

\paraf{(RQ5) Fix Strategy:} How can bugs be fixed? Identifying common bug fix strategies can guide the development of automated repair tools and inform developer guidelines.

\paraf{(RQ6) Temporal Evolution:} How do the distributions of bugs change over time?
Analyzing temporal evolution helps identify dynamic shifts in bug patterns, offering a detailed understanding of bugs across the development progress.
}
\medskip

To answer the RQs, we carefully select five representative LLM inference engines for our analysis, including Llama.cpp~\cite{llamacpp}, vLLM~\cite{vllm}, DeepSpeed~\cite{deepspeed}, Mlc-llm~\cite{mlcllm}, and TensorRT-llm~\cite{tensorrt-llm}.
These engines are commonly used in real-world LLM apps and exhibit diversity in their implementations, ensuring the validity and generalizability of our results.
We gather all the bug issues from the official repositories of these engines as of December 1, 2024.
Following the open coding procedure~\cite{seaman1999qualitative}, we review error reports, source code, developer discussions, and related pull requests (PRs) of issues to uncover the symptoms, root causes, and distribution of bugs, constructing a dataset with 929 manually labeled real-world bugs.

Based on the dataset, we identify six symptom types and construct a taxonomy of 28 specific root causes.
Our main findings include that 
(1) Testing LLM inference engines requires diverse oracles beyond mere crash detection, as over 35\% of bugs manifest as non-crash symptoms.
(2) Under undetermined LLM output, five factors (i.e., mis-encoded characters, incoherent semantics, inconsistent outputs, abnormal lengths, and repeated substrings) provide insights for oracle setting to detect non-crash symptom.
(3) We identify actionable diagnostic patterns for the observed bug symptoms. For instance, crashes, feature failures, and system hangs necessitate checks on functionality, environment, and configuration.
(4) Bug symptoms and root causes show a strong correlation across engines, with correlation coefficient greater than 0.5, indicating common bug patterns despite different engine implementations.
Based on our findings, we propose actionable implications for researchers, inference engine vendors, and LLM app developers, and offer general guidelines for developing
LLM inference engines

In summary, we make the following main contributions:
\begin{itemize}[left=0pt]
    \item To the best of our knowledge, we present the first empirical study on bugs in LLM inference engines.
    \item We present taxonomies of bug symptoms and root causes in LLM inference engines, which provide practical insights into bug detection and location. 
    \item We make our carefully collected dataset from our work public available for further research\footnote{\url{https://github.com/infbug/bugs-in-LLM-inference-engines}}.
\end{itemize}


%% file: sections/background.tex
\section{Background}
\label{sec:background}

In this section, we introduce the deployment workflow of LLMs and the main components of LLM inference engines.

\subsection{LLM Deployment}

\begin{figure}[t!]
    \centering
    \includegraphics[width=0.8\linewidth]{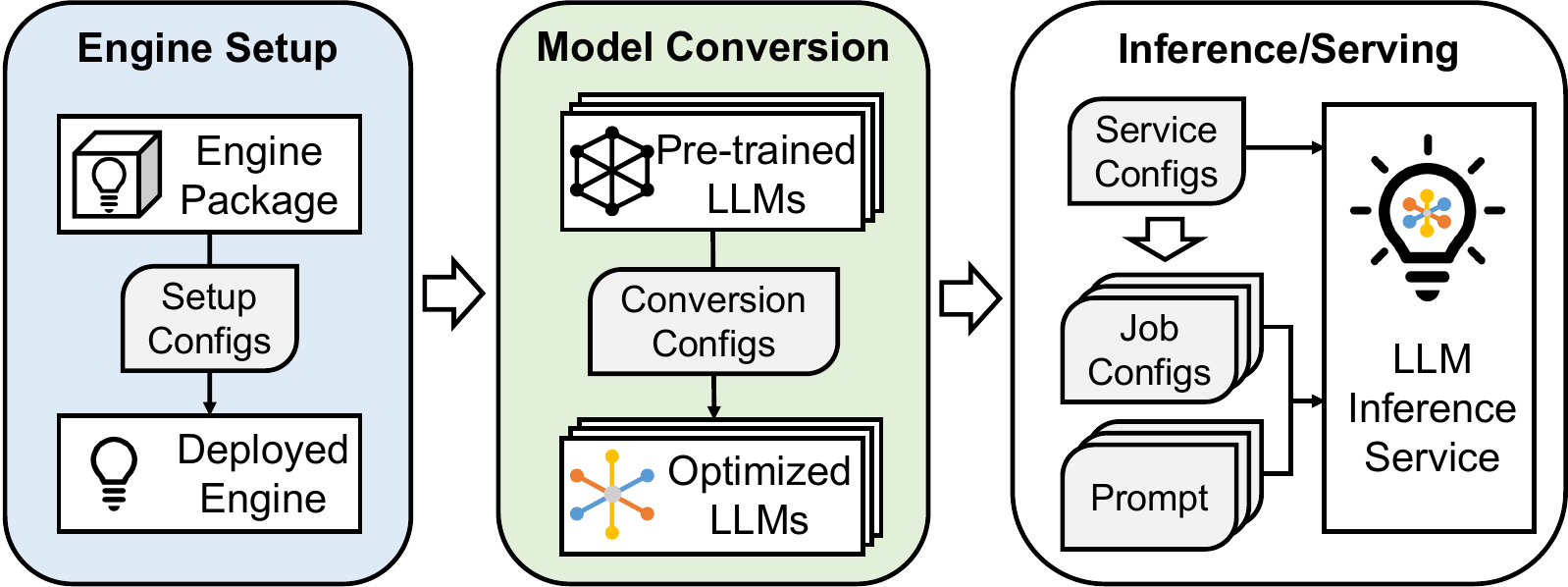}
    \caption{Workflow of LLM deployment.}
    \label{fig:workflow}
\end{figure}

LLM apps have gained widespread use, such as cloud-based chatbots~\cite{chatgpt}, coding assistants~\cite{corrabs240902977}, and personal AI assistants deployed on user devices~\cite{apple_intelligence,google_gemini}. 
In these apps, LLM inference engines are crucial components for enhancing inference efficiency and adapting LLMs to various environments.
Generally, the workflow of LLM deployment falls into three phases, including engine setup, model conversion, and inference/serving, as shown in Fig.~\ref{fig:workflow}.

Engine setup aims to deploy LLM inference engines onto target devices (e.g., distributed servers, PCs, and phones), either through source code compilation or pre-compiled binaries.
This phase involves installing necessary software modules and ensuring hardware compatibility with specific setup configurations.

Model conversion adapts models for specific engines, platforms, and devices through steps like format conversion, parameter optimization, and packaging. 
Pre-trained LLMs are adjusted for specific conversion configurations, and efficiency optimizations like quantization and sharding are applied based on the underlying environments. 
Given the high resource and performance demands of LLM inference, this process is crucial for ensuring the quality of LLM inference services.

Inference/serving aims to initialize services, load models, and process inference requests correctly and efficiently on target devices. 
Developers configure the LLM as a service, allowing users to set job-specific configurations (such as the model, backend, and inference strategies), submit prompts, and receive LLM responses.

\subsection{LLM Inference Engine}

\begin{figure}[t!]
    \centering
    \includegraphics[width=0.99\linewidth]{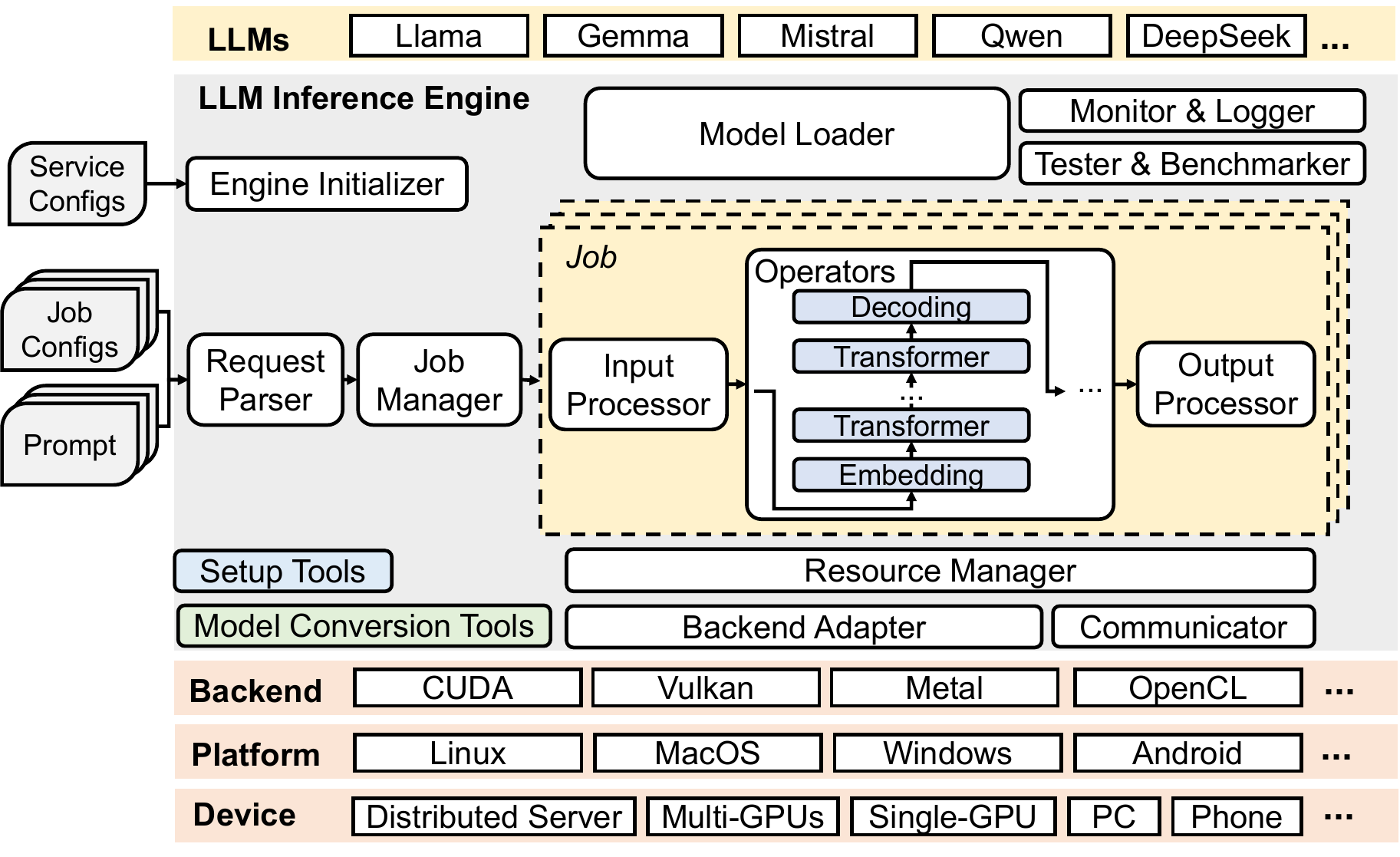}
    \caption{Architecture of LLM inference engine.}
    \label{fig:arch}
\end{figure}

\revise{
\parabf{Design goals.}
The primary design goals of LLM inference engines are compatibility and efficiency.
First, compatibility ensures that various series of LLM run smoothly across diverse environments. 
Engines like vLLM~\cite{vllm} and DeepSpeed~\cite{deepspeed} are tailored for diverse cloud-based distributed inference scenarios, while engines like Llama.cpp~\cite{llamacpp} and Mlc-llm~\cite{mlcllm} focus on compatibility with heterogeneous personal devices and backends.
Second, efficiency optimization focuses on integrating algorithms that maximize resource utilization and minimize inference latency under varying resource conditions. 
Engines generally achieve efficiency by optimizing model conversion, inference strategies, operator implementation, and resource management.

\parabf{Overall architecture.}
As shown in Fig.~\ref{fig:arch}, LLM inference engines act as a middleware layer that bridges hardware environments and LLMs. 
At the bottom level, diverse \emph{devices} (e.g., distributed servers, single-GPU servers, PCs, smartphones, and IoT devices) run on heterogeneous \emph{platforms} such as Linux, Windows, and Android, and rely on different \emph{backends} such as CUDA, Vulkan, and Metal. 
On top of these heterogeneous environments lies the \emph{LLM inference engine}, which abstracts the hardware complexity and provides unified support for loading and executing models. 
At the upper level, these engines enable dozens of popular LLM series, including Llama~\cite{llama} and Gemma~\cite{gemma}, to run seamlessly across devices and backends.}

\revise{
\parabf{Functionality.}
The functionality of LLM inference engines follows the deployment workflow of three phases. 
In the \emph{engine setup} phase, engines typically provide setup tools that allow custom construction and installation in the target environment. In the \emph{model conversion} phase, they supply conversion utilities to adapt pre-trained LLMs into formats optimized for specific backends and platforms. Finally, in the \emph{inference/serving} phase, users interact with LLMs through command-line interfaces or inference services. Upon receiving a start command, the engine initializes the service according to configurations, loads the converted model onto the backend, and processes user requests. For each request, the engine allocates resources, creates a job, checks the model, and performs auto-regressive token generation to produce outputs.}

\revise{
\parabf{Key components involving inference/serving.}
The inference/serving stage is the most complex functionality of LLM inference engines, relying on several key components, as illustrated in Fig.~\ref{fig:arch}.
At the beginning of inference/serving, the engine initializer processes start configurations to set up the inference environment, including loading dependent modules, configuring parameters, coordinating with the model loader to prepare LLMs, and opening the service port to await inference requests.
The model loader is responsible for loading the specified converted LLM onto the chosen backend.

For each inference request, the request parser processes the input prompts and job configurations, formatting them for the job manager. 
The job manager handles concurrent jobs and their contexts.
For each parsed request, the job manager creates a job from the configurations and collaborates with the model loader and resource manager to ensure smooth inference.
For each job, the input processor tokenizes the prompts and applies LLM-specific features, such as grammar constraints to enforce specific formats.
The engine then generates tokens auto-regressively based on backend-specific operators. 
Finally, the output processor detokenizes the output tokens, converts them to the specified output format, and sends the response to the user.

The resource manager employs optimized algorithms to manage critical computational and memory resources for LLM inference, such as the key-value cache (KV cache), which stores and reuses the generated key and value vectors to reduce redundant calculations.
The backend adapter manages API interactions with the hardware, supporting the resource manager's operations.
The communicator is tailored for distributed environments, facilitating message passing and status synchronization between servers.
Engines also typically include monitors and loggers to track metrics and logs during inference, as well as testing and benchmarking tools to evaluate inference accuracy and performance.
}

%% file: sections/methodology.tex
\section{Methodology}
\label{sec:methodology}
\begin{figure}[t]
    \centering
    \includegraphics[width=0.8\linewidth]{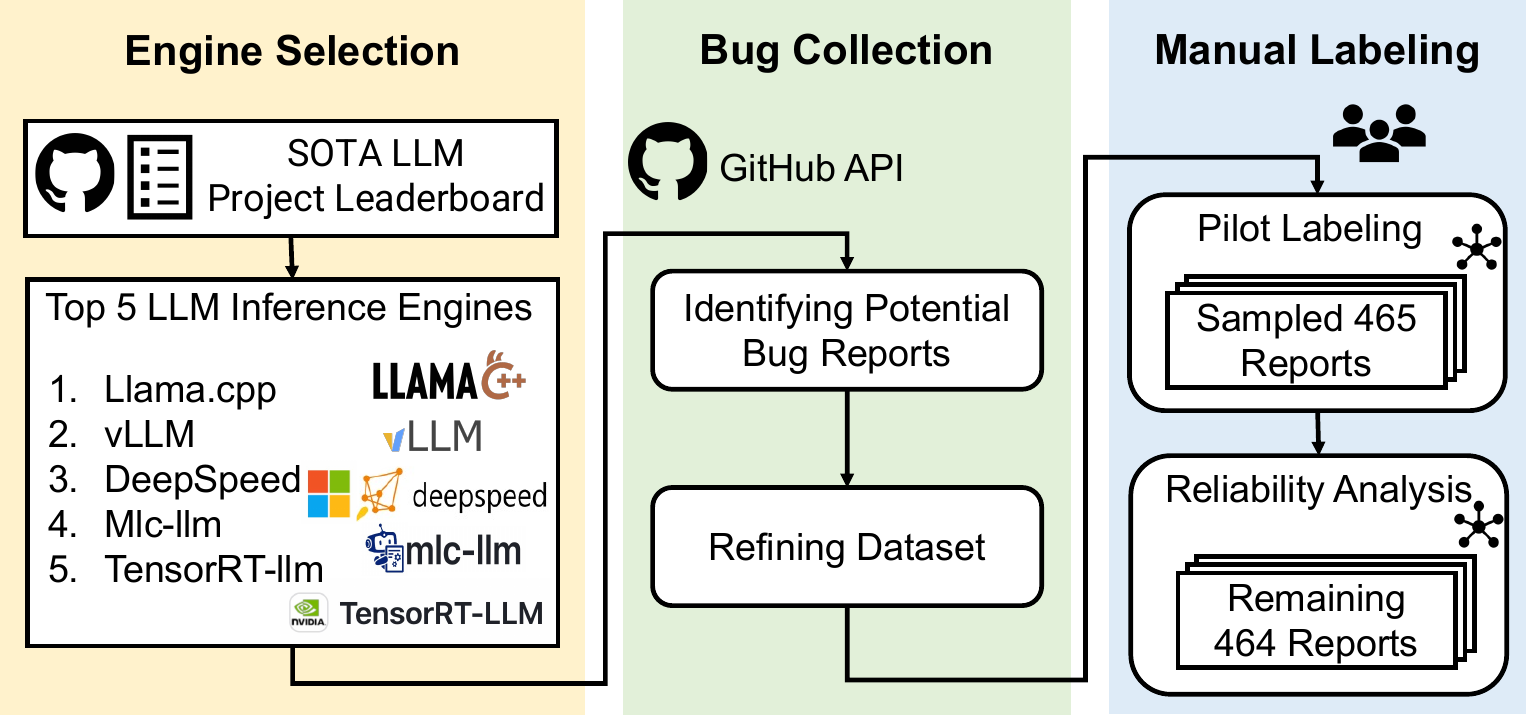}
    \caption{Overview of methodology.}
    \label{fig:method}
\end{figure}

Fig.~\ref{fig:method} shows an overview of our methodology. We begin by selecting representative LLM inference engines, then collect relevant bugs, and finally perform manual labeling to characterize them.


\subsection{Engine Selection}
With the rising popularity of LLMs, numerous LLM inference engines have been developed, making it impractical to consider them all. 
Therefore, we select representative ones for our study. 
Notably, we focus on LLM-specific inference engines, excluding those designed for general DL model inference.

We begin with a widely recognized leaderboard of state-of-the-art LLM training and inference projects actively maintained by the community~\cite{awesomellm}. 
Since our study focuses on inference-related bugs, we first exclude engines dedicated to LLM training. We also remove closed-source engines due to their lack of accessible bug reports. From the remaining candidates, we select the top five LLM inference engines based on the number of stars in their official GitHub repositories, a widely used indicator of popularity~\cite{liu2024research,li2020exploratory}.
As a result, we adopt Llama.cpp~\cite{llamacpp}, vLLM~\cite{vllm}, DeepSpeed~\cite{deepspeed}, Mlc-llm~\cite{mlcllm}, and TensorRT-LLM~\cite{tensorrt-llm}. The details of these engines are summarized in Table~\ref{tab:select_engine}.

\input{tables/select_engine}

These engines are representative for several reasons. First, they have received over 10,000 GitHub stars, indicating strong popularity. Second, they have been adopted by major organizations. For example, vLLM serves as the backend for Hugging Face, DeepSpeed is integrated into Microsoft’s infrastructure, and TensorRT-LLM is utilized by Nvidia. Third, they show diversity in targeted environments. Llama.cpp and Mlc-llm are designed for universal deployment, supporting various operating systems (such as Linux, Windows, macOS) and platforms, including PCs, mobile devices, and IoT devices. vLLM and DeepSpeed are optimized for cloud environments, while TensorRT-LLM supports both cloud-based deployments and edge servers. Finally, these engines incorporate cutting-edge features in LLM inference, such as continuous batching, paged attention, speculative decoding, and machine learning compilation.

\subsection{Bug Collection}
To collect bugs from the selected LLM inference engines, we focus on their official GitHub repositories. Specifically, we first follow previous work~\cite{chen2023toward,zhang2018empirical,LiuGCWZMWJ23} to leverage the GitHub issue tracking system to gather potential bug reports for each engine. We then refine the collected data and exclude invalid reports.

\subsubsection{Identifying Potential Bug Reports.}
Since not all issues are related to bugs, we follow previous work~\cite{chen2021empirical} by using repository-specific labels to identify potential bug reports. We thoroughly review each repository to understand its labeling system. For Llama.cpp, vLLM, Mlc-llm, and TensorRT-llm, `label:bug' indicates bug reports. For DeepSpeed, which supports both training and inference, bug reports related to inference are identified only when both `label:bug' and `label:inference' are used together. We then filter the issues based on these labels for each engine. Additionally, following prior studies~\cite{chen2021empirical}, we exclude open issues, as they are typically unresolved and may lack clear insights into the root causes. After this process, we obtain 184, 1200, 187, 551, and 366 potential bug reports for Llama.cpp, vLLM, DeepSpeed, Mlc-llm, and TensorRT-llm, respectively, as of December 1, 2024.

\subsubsection{Refining Dataset}
\revise{
We observe false positives in the collected bug reports, as the bug-related tags can be misapplied.
To address this, manual filtering is necessary to remove these false positives. However, manually analyzing all the issues is a challenging task. 
Therefore, we adopt a sampling approach following previous work~\cite{zhang2019empirical,chen2020comprehensive,LouCCHZ20}, randomly selecting a subset of the bug reports while ensuring a 95\% confidence level and a 5\% confidence interval. 
This sampling process is applied separately to each engine, resulting in 125, 292, 126, 227, and 188 bug reports for Llama.cpp, vLLM, DeepSpeed, Mlc-llm, and TensorRT-llm, respectively.

After sampling, the first and second authors, each with over two years of experience in LLM-based application development, serve as inspectors and separately label false positives.
In this process, they review the title, description, code snippets, comments, answers, and labels of each issue to determine whether the issue meets our exclusion criteria. 
The applied exclusion criteria include:

\begin{itemize}[left=0pt]
    \item \textbf{Not a bug.} Issues that only contain feature requests or general inquiries are excluded, as they do not represent bugs.
    \item \textbf{Not a bug of the engine itself.} Issues arising from user misunderstandings or misuse of the engines are excluded, as they do not reflect inherent engine bugs.
    \item \textbf{No informative comments.} Issues without substantial or informative comments are excluded, as they do not provide sufficient details for analysis.
\end{itemize}

The exclusion criteria are straightforward to identify. For instance, issues containing only the initial reports and lacking any responses can be easily identified as 'No informative comments.' 
Consequently, the two inspectors encountered no conflict in identifying false positives due to the clear criteria. 
Furthermore, because the issues were randomly sampled, the absolute number of the excluded issues was relatively small. 
Specifically, they excluded 1, 2, 5, 13, and 8 false bug reports from Llama.cpp, vLLM, DeepSpeed, Mlc-llm, and TensorRT-llm, respectively. 
After this manual refinement, we are left with a dataset of 929 real-world bugs related to LLM inference engines. The final counts are 124, 290, 121, 214, and 180 reports for Llama.cpp, vLLM, DeepSpeed, Mlc-llm, and TensorRT-llm, respectively.}

\revise{
For each issue in the refined dataset, we retrieved the creation and closure times using the GitHub API~\cite{github_api} to calculate the duration required for bug repair.
}

\subsection{Manual Labeling}

We manually analyze the 929 bugs to identify their symptoms and root causes, constructing corresponding taxonomies for LLM inference engine bugs. 
Following previous work~\cite{chen2021empirical,ZhangCWCLMMHL24,sigsoftWenCLL00JL21,msrWangCZ23}, the manual analysis involves two key steps: initially constructing the taxonomies through a pilot labeling process, followed by a reliability analysis to evaluate the taxonomies. The entire manual analysis takes over 500 man-hours.

\subsubsection{Pilot Labeling}
We randomly select 50\% of the dataset, totaling 465 issues, for pilot labeling. As the first study to characterize bugs in LLM inference engines, we develop the taxonomies for bug symptoms and root causes from scratch, following the open coding procedure~\cite{seaman1999qualitative}.

First, two authors, each with over two years of experience in LLM-based application development, act as inspectors and independently generate initial codes for bug symptoms and root causes.
To fully understand the content of each issue, they review the title, description, code snippets, comments, answers, labels, associated PRs, and any referenced URLs. They then assign brief phrases as initial codes to capture the key features of each issue.

For symptoms, the inspectors focus on what the bug reporter observes and at which stage of LLM deployment the bug occurs. 
This information is often found in the issue description, call stacks, and related reproduction steps.
For example, a reporter states that `\emph{loading a model that has been quantized fails}' in issue \#8116 of vLLM. 
From the perspective of vLLM, the model loading is one of the steps in inferece/serving stage.
\rerevise{Because the engine terminated unexpectedly with an unhandled exception and a stack trace during model loading (as shown in \autoref{bug:vllm-8116}), preventing it from serving requests, the inspectors classify this case as a crash occurring in the inference/serving stage.}

\begin{lstlisting}[
  % language=verbatim, 
  caption={\rerevise{Error messaegs in vLLM issue \#8116.}},
  label={bug:vllm-8116}
]
File "xxx/gpt_bigcode.py", line 356, in load_weights
    weight_loader(param, loaded_weight)
File "xxx/linear.py", line 779, in weight_loader_v2
    self._load_fused_module_from_checkpoint(param, loaded_weight)
File "xxx/linear.py", line 762, in _load_fused_module_from_checkpoint
    loaded_weight_shard = loaded_weight.narrow(param.output_dim,
    ...
\end{lstlisting}

\revise{
In contrast, identifying root causes is more challenging and often requires in-depth analysis. Inspectors determine how the code failed by examining call stacks, bug reproduction experiments, confirmed comments from developers, issue descriptions, and associated PRs. 
For instance, in issue \#3821 of vLLM, the associated PR \#4021 states `\emph{avoid allocating a CUDA context for every GPU}', and modifies the allocation of the cache dictionary in its commits.
Based on this, inspectors infer that the bug originated from incorrect resource allocation.
If a bug involves multiple root causes, inspectors assign it to all relevant root cause categories.
Additionally, to investigate the vulnerable components in LLM inference engines, they also mark the associated component for each bug.
This is primarily identified by the associated modular code components in the LLM inference engine.
If the relevant faulty code is not within a modular component, they infer its component based on its functionality. 
For instance, the issue \#3821 of vLLM is assigned to the resource manager component.
Multi-component bugs are assigned to all relevant components.}

Then, inspectors discuss to achieve agreements on the initial codes.
In case of disagreements, another author, who has over two years of LLM inference engine deployment experience, acts as an arbitrator and joins the discussion to help reach consensus. 
After generating the initial codes for issues, inspectors group similar codes into categories.
The grouping process is iterative, with inspectors continuously reviewing the categories and issues to refine the taxonomies.
Bugs associated with multiple symptoms or root causes are assigned to all relevant categories. 
Finally, the two inspectors and arbitrator approve all categories in the symptom taxonomy and the root cause taxonomy.

\subsubsection{Reliability Analysis}
This step aims to validate the reliability of the pilot taxonomies using the remaining 464 issues.

The two inspectors independently classify each issue according to the symptom and root cause taxonomies, following the coding scheme established during pilot labeling. If an inspector is unsure how to classify an issue, it is marked as `Pending.'

To evaluate the reliability of our coding schema and procedure, we measure the inter-rater agreement between the two inspectors using Cohen's Kappa ($\kappa$)~\cite{cohen1960coefficient}, a widely adopted metric in empirical software engineering studies~\cite{chen2021empirical,chen2023toward}. The computed $\kappa$ values for symptom and root cause labeling are 0.927 and 0.873, respectively, indicating almost perfect agreement and confirming the reliability of our coding schema.

The arbitrator participates in discussions to resolve any labeling conflicts between the inspectors. For issues marked as `Pending,' they collectively determine whether new categories are needed. 
Ultimately, no new category of symptoms and root causes is introduced. 
All pending issues, comprising only 0.9\% of the total analyzed cases, are successfully classified within the existing taxonomies, demonstrating the saturation of the taxonomies. The final labeling results are reviewed and approved by all participants.

\revise{
\subsubsection{Fix Strategy Analysis}

Based on the constructed root cause taxonomy, we extend our study to include bug fix strategies for the root causes. 
The methodology follows the same rigorous procedure as the symptom and root cause labeling.

Utilizing the pilot labeling dataset mentioned earlier, we employ the GitHub API~\cite{github_api} to retrieve the associated fix commits for bugs linked to each internal root cause category (see Section \ref{sec:rq2}).
Then, the two inspectors independently read the contents of the retrieved fix commits, including the title, developer conversations, and changed files.
Next, they assign initial fix strategies to each fix commit based on the contents.
They then engage in discussions to reach agreements on the initial codes.
The arbitrator is involved to address any disagreement.
Subsequently, the three authors iteratively refine the initial codes by grouping similar codes and continuously reviewing the categories and contents of the fix commits.
Finally, they obtain initial fix strategies for bugs associated with each internal root cause category.

For reliability analysis, following the same procedure, two inspectors label the fix commits of remaining issues based on the initial codes of fix strategies.
The $\kappa$ value for the fix strategy labeling is $0.8179$, indicating almost perfect agreement.
No new category of fix strategies is introduced.
The categories that fix at least 5 issues are identified as common fix strategies.
}

%% file: tables/select_engine.tex

    




\begin{table}[t!]
\centering
\caption{Details of selected LLM inference engines (Data as of \textit{Oct. 2025}). ``GGML'' is the underlying tensor library of  Llama.cpp that enables the efficient execution of LLMs on consumer-grade hardware. ``Universal'' refers to the engine's capability for broad deployment across heterogeneous operating systems and hardware.}
\label{tab:select_engine}
\resizebox{\textwidth}{!}{%
\begin{tabular}{lll|rr|rrrr}
\toprule
\multirow{2}{*}{\textbf{Engine}} & 
\multirow{2}{*}{\textbf{Provider}} & 
\multirow{2}{*}{\textbf{Target Env.}} & 
\multicolumn{2}{c|}{\textbf{GitHub Repo}} & 
\multicolumn{4}{c}{\textbf{Codebase Size (LoC)}} \\
\cmidrule{4-5} \cmidrule{6-9}
&&& 
\textbf{Stars} & 
\textbf{Commits} & 
\textbf{Total} &
\textbf{C/C++} & 
\textbf{Python} & 
\textbf{CUDA} \\
\midrule
Llama.cpp & GGML Team & Universal & 87,810 & 6,765 & 230,397 & 194,360 & 18,576 & 17,461 \\
vLLM & UCB & Cloud & 60,137 & 10,462 & 458,618 & 7,955 & 413,545 & 37,118 \\
DeepSpeed & Microsoft & Cloud & 40,391 & 2,956 & 121,965 & 9,460 & 98,246 & 14,259 \\
Mlc-llm & CMU & Universal & 21,473 & 1,721 & 1,508,976 & 422,362 & 473,255 & 613,359 \\
TensorRT-llm & Nvidia & Cloud/Edge & 11,856 & 3,173 & 20,754,106 & 20,296,665 & 385,877 & 86,564 \\
\bottomrule
\end{tabular}
}
\end{table}


%% file: sections/symptom.tex
\section{RQ1: Symptoms}
\label{sec:rq1}

To clarify the impact of bugs and provide insights for test oracle, we explore the bug symptoms and their distribution across LLM inference engines.
%

\revise{
\subsection{Symptom Classification}

We classify the observed symptoms into six unique categories, as shown in Table~\ref{tab:symptom_description}.
\emph{Crashes (S1)} refer to unexpected engine termination without recovery capability, such as the unrecoverable termination of compilation or inference process due to out-of-bound memory access (Llama.cpp issue \#10176).
\emph{Unexpected Output (S2)} denotes the generation of incorrect model parameters, garbled tokens, or responses that are illogical or irrelevant to the prompt.
An example of this is the generation of a random mix of words in different languages, as reported in Llama.cpp issue \#1735.
\emph{Feature Failure (S3)} refers to the engine remaining running but failing to perform specific features, such as the invalid aliases configured for the model instance noted in vLLM issue \#8947.
\emph{Abnormal Performance (S4)} refers to token generation with unusually high or low resource consumption or latency, like the absence of GPU consumption and increased latency in Llama.cpp issue \#3806.
\emph{System Hang (S5)} indicates that the engine becomes unresponsive but does not terminate, such as when the model quantization process stalls without completion, as seen in TensorRT-llm issue \#1356.
\emph{Silent Error (S6)} involves internal anomalies that evade detection mechanisms. 
For example, in the Llama.cpp issue \#9245, the engine produced corrupted model files without generating any warnings.
}

\finding{Six bug symptoms are identified, including crash, unexpected output, feature failure, abnormal performance, system hang, and silent error.}{finding:symptoms}

\revise{
\subsection{Symptom Distribution}
We study the distribution of identified symptoms across the three phases of LLM inference engine deployment workflow (i.e., engine setup, model conversion, and inference/serving).
As shown in Table~\ref{tab:symtoms}, these symptoms vary in distribution across these phases.
Note that although different phases may exhibit the same symptom (e.g., a crash), their impacts could differ. 
For instance, a crash during the setup phase may prevent an engine from being installed in the specific environment. Meanwhile, a crash during model conversion results in a faulty model, rendering it unusable during the inference/serving stage. Additionally, a crash during inference/serving stage disrupts user interaction and service availability.}

\subsubsection{Engine Setup}

During engine setup, we identify 136 bugs (15\% of total), covering three symptoms including \emph{S1}, \emph{S5}, and \emph{S2}.

Specifically, most (97\%) symptoms in this phase are \emph{Crash (S1)} with 132 instances, indicating abrupt termination during engine source code compilation or pre-compiled binary execution.
For example, in Llama.cpp issue \#279, the engine compilation process terminated unexpectedly due to missing Apple Accelerate header files. 
Among the remaining four bugs, three exhibit \emph{System Hang (S5)}, where the process becomes unresponsive without crashing.
For example, in vLLM issue \#3601, the construction of the docker image was stuck for a very long time.
The last remaining bug falls into \emph{Unexpected Output (S2)}, where an incorrect warning occurred during the engine construction in DeepSpeed issue \#6486.

\finding{Most (97\%) bugs in engine setup phase lead to a crash.}{finding:sym_es}

\begin{table}[t!]
\caption{Symptom Categories in LLM Inference Engines.}
\label{tab:symptom_description}
\vspace{-1em}
\begin{center}
\resizebox{0.95\textwidth}{!}{
    \begin{tabular}{@{}l>{\raggedright\arraybackslash}p{2cm}>{\raggedright\arraybackslash}p{12cm}@{}}
    \toprule
    \textbf{ID} & \textbf{Symptom} & \textbf{Description} \\
    \midrule
    S1 & Crash & Engines terminate unexpectedly without recovery capability. \\
    S2 & Unexpected Output & Engines output incorrect model parameters or tokens that are unreadable, incoherent, unrelated to prompts, or fail to maintain logical capability. \\
    S3 & Feature Failure & Engines remain operational but fail to perform specific features. \\
    S4 & Abnormal Performance & Engines generate tokens with excessive or insufficient resource usage, or exhibit abnormal generation latency. \\
    S5 & System Hang & Engines remain operational but become unresponsive.\\
    S6 & Silent Error & Engines experience internal anomalies without error detection. \\
    \bottomrule
    \end{tabular}
}
\end{center}
\end{table}

\begin{table}[t!]
\caption{Number of bug symptoms in different phases.}
\label{tab:symtoms}
\vspace{-1em}
\begin{center}
\resizebox{0.6\linewidth}{!}{
    \begin{tabular}{l|rrrrrr|r}
    \toprule  
    & \textbf{S1} & \textbf{S2} & \textbf{S3} & \textbf{S4} & \textbf{S5} & \textbf{S6} & \textbf{Total}\\
    \midrule  
    \textbf{Engine Setup} & 132 & 1 & 0 & 0 & 3 & 0 & 136 \\
    \textbf{Model Conversion} & 108 & 0 & 14 & 0 & 2 & 8 & 132 \\
    \textbf{Inference/Serving} & 363 & 121 & 93 & 53 & 30 & 4 & 664 \\
    \cmidrule(lr){1-8} \textbf{Total}  & 603 & 122 & 107 & 53 & 35 & 12 & 932 \\
    \bottomrule
    \end{tabular}
}
\end{center}
\end{table}

\subsubsection{Model Conversion}

We identify 132 bugs (14\% of total) in the model conversion phase, covering four symptoms including \emph{S1}, \emph{S5}, \emph{S3}, and \emph{S6}.

Specifically, the majority (82\%) symptoms are classified into \emph{Crash (S1)} with 108 instances, where the model conversion process terminates abruptly.
For instance, in TensorRT-llm Issue \#899, the conversion of LLM to `SmoothQuant' format failed due to mismatched shape of tensors.
Two bugs show \emph{System Hang (S5)}, where the model quantization process stalls without termination (e.g. TensorRT-llm issue \#1356).

22 bugs result in the generation of incorrect model files.
Among these bugs, 14 are classified as \emph{Feature Failure (S3)}, where the generated complete model files are incomplete or misconfigured under specific settings.
For example, in TensorRT-LLM issue \#1245, the conversion tools failed to generate complete `INT4' model file under the AWQ quantization settings.
The other 8 bugs are identified as \emph{Silent Error (S6)}, where conversion tools generate seemingly complete yet incorrect model files without any warnings.
For example, in Llama.cpp issue \#9245, although the input model file was broken, the engine generated model files without warnings.
This highlights the need for thorough checks of generated model files to prevent incorrect models from causing severe inference errors.

\finding{In model conversion phase, 83\% of bugs prevent the generation of model files (\emph{S1}, \emph{S5}) while 17\% lead to the creation of incorrect model files (\emph{S3}, \emph{S6}).}{finding:sym_mc}

\subsubsection{Inference/Serving}
We observe 664 bugs (71\% of total) in the inference/serving phase, covering all six identified symptoms.
\revise{We describe them in a gradual order, beginning with the most disruptive failures, moving to cases where the engine continues to run but behaves incorrectly, and concluding with other symptoms observed when outputs are still produced.}

\revise{\parabf{Service-disrupting bugs}. Most (59\%) bugs cause the engine to stop functioning, involving symptoms of \emph{S1} and \emph{S5}.}
\emph{Crash (S1)} is the most common symptom, with 363 instances involving initialization failures or irrecoverable service termination. 
For example, in vLLM issue \#8952, the server crashed when receiving requests with configurations like `response\_format' or `guided\_json'.
\emph{System Hang (S5)} occurs in 30 cases, where the service becomes unresponsive without crashing, as seen in Mlc-llm issue \#306, where API calls hang during prefill function execution.

\revise{\parabf{Functional anomalies bugs.} 93 bugs make engines exhibit functional anomalies while remaining operational, which are classified into \emph{Feature Failure (S3)}.}
\rerevise{For instance, in vLLM issue \#8947, the engine's argument parser incorrectly required the alias (`\texttt{served-model-name}') to be the first parameter. 
A configuration where the alias appeared later, such as `\texttt{(host=`xxx', port: `xxx', served-model-name=`MyModel')}', would cause the engine to erroneously ignore the alias (\texttt{served-model-name=`MyModel'}) silently.
In this case, vLLM performed model inference without setting the configured alias. 
Therefore, this was classified as a Feature Failure.
}

\revise{\parabf{Bugs during output generation.}
Bugs can also happen when engines start generating output tokens, exhibiting as \emph{S2}, \emph{S4}, and \emph{S5}.}
\revise{\emph{Unexpected Output (S2)}, with 121 instances, occurs when the service produces user-unexpected tokens or metric values (e.g., probability distribution). 
This category consists of two subtypes: (i) unexpected tokens from LLM generation and (ii) unexpected non-token outputs. Unexpected tokens refer to cases where the LLM-generated tokens are not human-readable, semantically coherent, logically consistent, or prompt-relevant, such as producing gibberish or irrelevant text. 
\rerevise{For example, Llama.cpp issue \#1735 reported unreadable results (e.g., ` \texttt{\&amp Vertigowebsitesearch engines Search engines Search}'), which were classified as unexpected tokens. }
The second subtype includes non-token outputs, which occur when components like setup tools, model conversion tools, and monitoring tools within the engines generate erroneous setup files, model files, model parameter values, or metric values. 
For example, Llama.cpp issue \#10285 reported a case of incorrect output of model parameter and size counts due to incorrect algorithm implementation.}
Showing parameter size consistency with PyTorch but logits deviations
\emph{Abnormal Performance (S4)}, with 53 bugs, manifests as either excessively high or low performance metrics (e.g., latency, throughput, and resource utilization).
\rerevise{For instance, issue \#2488 of DeepSpeed reports a case of abnormal performance, where the supposedly optimized DeepSpeed model was slower than the vanilla model, with its average latency increasing from \textasciitilde8566ms to \textasciitilde8811ms.}
\emph{Silent Error (S6)}, with 4 bugs, involves hidden code defects that may disrupt computations.
For example, the issue \#6030 of vLLM reports that an if-else condition branch contained identical code blocks.


\finding{
The inference/serving phase has the highest bug proportion (71\%) and exhibits all six symptom types.
}{finding:sym_phases}

\begin{figure}[t!]
    \centering
    \includegraphics[width=0.8\linewidth]{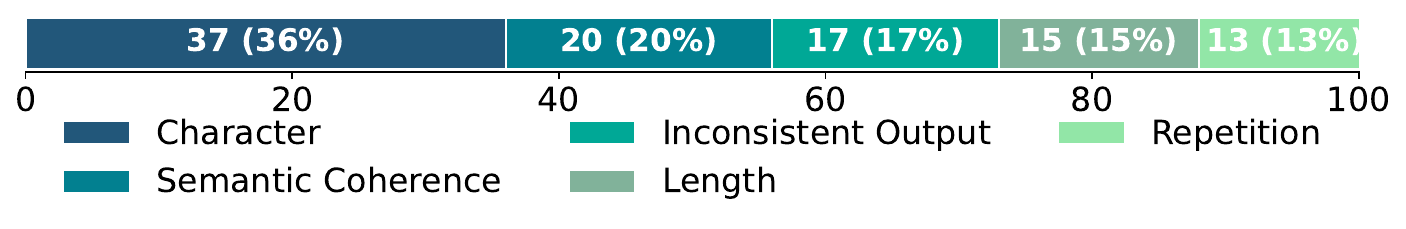}
    \caption{Distribution of unexpected factors.}
    \label{fig:distribution_of_unexpected_tokens}
\end{figure}

Note that a key challenge in testing LLM inference is setting test oracles for their inherently undetermined output tokens. 
Therefore, we further analyze bugs related to unexpected output tokens to determine what constitutes `unexpected' outputs.

Following the labeling process in section~\ref{sec:methodology}, among the 121 `Unexpected Output' bugs in inference/serving phase, we further identify 102 bugs
\footnote{\revise{The remaining 19 cases involve unexpected non-token outputs.}}
with unexpected output tokens, covering 5 factors, whose distribution is shown in Fig.~\ref{fig:distribution_of_unexpected_tokens}.
\revise{
Among the unexpected token output, the most (36\%) observed factor is unexpected \emph{Character (UT1)}, with 37 bugs, which indicates that the output tokens of LLMs involve massive garbled text (e.g. Llama.cpp issue \#1735).
\emph{Semantic Coherence (UT2)} is the second most (20\%) used factor, found in 20 cases, implying that the outputs of LLM are meaningless (e.g. Llama.cpp issue \#3782).
\emph{Inconsistent Output (UT3)} indicates that even after disabling all randomness settings, the LLM outputs still vary across different optimization algorithms or runtime environments.
This factor accounts for 17\%, with 17 instances.
For instance, in vLLM issue \#590, output tokens differ between the original model and the model served using vLLM.
15\% of bugs, found in 15 issues, are identified by unexpected \emph{Length (UT4)} of output tokens, occur when outputs are unusually short or excessively long, never stopping, exemplified by vLLM issue \#9448.
\emph{Repetition (UT5)} issues, seen in 13 cases, involve excessive duplication of strings.
For example, in DeepSpeed issue \#4903, the generated tokens abnormally repeat the `e' character.
}

\finding{
Despite non-deterministic outputs of LLMs, five factors (i.e. character, semantic coherence, inconsistent output, length, and repetition) provide insights for oracle settings.
}{finding:token_oracle}

Overall, \emph{Crash (S1)} is the most prevalent (65\%) symptom across all three phases. 
\emph{Unexpected Output (S2)} ranks second at 13\%, primarily occurring during the inference/serving phase. 
\emph{Feature Failure (S3)} follows at 11\%, with \emph{Abnormal Performance (S4)} at 6\%, \emph{System Hang (S5)} at 4\%, and \emph{Silent Error (S6)} at 1\%.
Among these symptoms, crashes are commonly used for bug detection in DL libraries~\cite{deng2023large,deng2024large} due to their distinct characteristics and easy detection through core dumps or stack traces.
Moreover, unexpected outputs are the second most common symptoms, comprising nearly 20\% during the inference/serving phase.
This underscores the need for suitable test oracles for LLM outputs and highlights the significance of Finding~\ref{finding:token_oracle}.

\finding{
While crashes are the most common, over 35\% of bugs show other symptoms,  necessitating diverse oracles beyond crashes for LLM inference engine testing.
}{finding:sym_distribution}

%% file: sections/root_cause.tex
\section{RQ2: Root Causes}
\label{sec:rq2}
To identify the vulnerabilities and obtain insights into bug detection and localization, we study the root causes and their corresponding components of LLM inference engines.

\subsection{Root Cause Classification}

\begin{figure*}[t!]
    \centering
    \includegraphics[width=\linewidth]{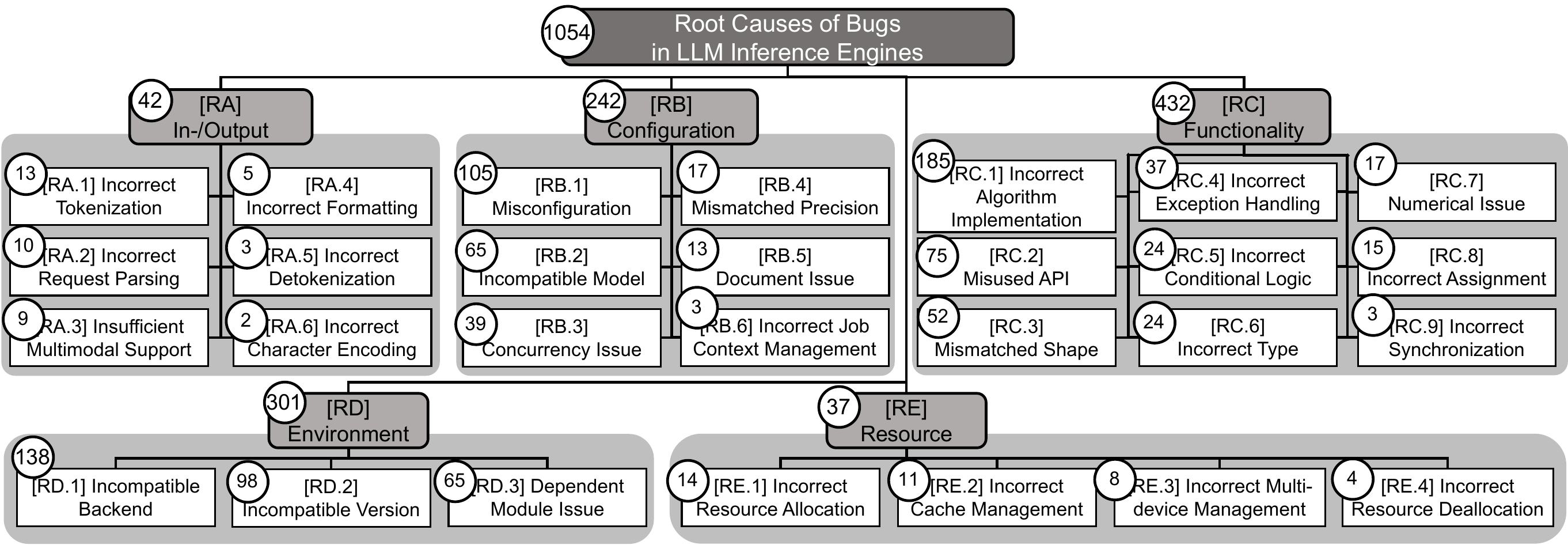}
    \caption{Taxonomy of root causes of bugs in LLM inference engines. The top-left circle numbers indicate bug counts.}
    \label{fig:root_causes}
\end{figure*}

Following the labeling process in section~\ref{sec:methodology}, we identify 28 specific leaf categories of root causes (e.g., \emph{Incompatible Model}) and group them into 5 inner categories linked to key factors in LLM inference process (including \emph{In-/Output}, \emph{Configuration}, \emph{Algorithm}, \emph{Environment}, and \emph{Resource}).
Fig.~\ref{fig:root_causes} shows our hierarchical taxonomy.
For each category, the top-left circle number indicates the bug count it causes.
We discuss and exemplify each inner category below.


\finding{
We develop a taxonomy of 28 specific root causes of bugs in LLM inference engines, indicating the diversity of bug sources in these complex engines.
}{finding:rc_taxonomy}

\subsubsection{[RA] In-/Output}

In-/output processing is crucial for user interaction during LLM inference. 
The engine receives and parses user input (including job configurations and prompts), decodes characters, and performs tokenization to prepare for inference. 
After inference, it decodes the generated tokens, formats the output, and returns it. 
Engines support various user interaction methods, such as HTTP requests and the command line. 
We observe 41 cases originating from in-/output processing, accounting for 4\% of cases, spanning six leaf categories.
Next, we detail each root cause following the processing flow.

For user input, inference engines must parse the formatted job configurations and prompts.
The \emph{Incorrect Request Parsing (RA.2)}, with 10 cases, indicates a failure to validate and parse requests, or provide feedback for malformed inputs.
For example, in Llama.cpp issue \#3766, a request with malformed `\emph{system\_prompt}' configuration caused a system hang without warning due to inadequate request validation.
Additionally, in this process, engines must avoid \emph{Incorrect Character Encoding (RA.6)}, with 2 cases, to accurately encode the characters of prompts.
RA.6 highlights the mishandling of incompatibilities between the engine's and users' character encodings.
\rerevise{For instance, in Mlc-llm issue \#804, the engine failed to correctly output certain characters (such as emoji and CJK characters).}

After parsing requests, engines will tokenize the prompt for subsequent inference. 
\emph{Incorrect Tokenization (RA.1)}, with 12 cases, indicates the use of misconfigured tokenizer or incorrect tokenization algorithm, leading to wrong token splits.
\rerevise{For instance, in Llama.cpp issue \#290, the tokenizer completely failed to process the input string \texttt{`\ china'} (a space `\texttt{\ }' and a word `\texttt{china}'). It reported only one token corresponding to the space (`\texttt{\ }') in the prompt while ignoring the actual word (`\texttt{china}'), leading to an empty effective input for the model.}
RA.1 is the primary cause of in-/output relevant bugs, underscoring the need for meticulous attention to tokenizer checks, handling, and error tolerance.
Moreover, engines face challenges in parsing multimodal data, often encountering issues with format parsing and dimension alignment for multimodal inputs, i.e., \emph{Insufficient Multimodal Support (RA.3)} with 9 cases. 
For example, in vLLM issue \#9128, the engine failed to correctly merge image and text embeddings.

After token generation, abnormal outputs can occur if the output pipeline misconfigures detokenization settings or algorithms, i.e., \emph{Incorrect Detokenization (RA.5)} with 3 cases.
For example, in Llama.cpp issue \#4036, the output Unicode characters displayed correctly in the command line but failed in server mode.
\emph{Incorrect Formatting (RA.4)} of user-specified output, with 5 cases, can also lead to bugs, as demonstrated in vLLM issue \#5334, where the requested log probability of tokens was not returned.

\finding{
Although in-/output relevant bugs are less frequent (comprising 4\% of bugs), they significantly impact user interaction, necessitating careful checking, processing, and error handling of character encoding, tokenizer, and user-specific formatting.
}{finding:rc_inoutput}

\subsubsection{[RB] Configuration}

Configurations are crucial throughout the entire LLM deployment workflow. 
Engine setup requires configurations such as hardware acceleration support.
Then, model conversion requires settings such as precision. 
Next, inference/serving involves service configurations to initiate LLM services, specifying factors like the path of LLMs. 
After that, users can also specify job configurations, such as the maximum context length. 
We identify 242 cases (23\% of total) caused by configuration issues, divided into six categories. 
This vast number of bugs highlights the challenges in configuration setup and management in LLM inference.
Next, we discuss the detailed root causes linked to configurations.


\emph{Misconfiguration (RB.1)} indicates the incorrect setup or implementation of individual configurations or their combinations.
For instance, in DeepSpeed issue \#3628, the engine failed to split the LLM under configurations of tensor parallelism. 
This category, with 105 cases, is highly bug-prone due to the engine's extensive parameter support.
This underscores the necessity of rigorous testing to ensure each configuration's correctness and compatibility with others.

After correct configuration setup, 82 cases are linked to the configured LLMs, including \emph{(RB.2)} and \emph{(RB.4)}.
65 cases indicate incorrect configuration and loading of LLMs, referred to as \emph{Incompatible Model (RB.2)}.
For example, in TensorRT-llm issue \#1634, quantization of Llama-3-70B-Instruct with w4a16 parameters failed because the engine lacked support for the Llama-3 architecture. 
Engines require individual adaptations for each LLM, which can lead to potential oversights for new models.
This highlights the necessity for comprehensive engine compatibility testing for new LLMs.
Additionally, 17 cases exhibit incorrect precision settings impacting model conversion, loading, and inference, referred to as \emph{Mismatched Precision (RB.4)}.
For instance, in TensorRT-llm issue \#1303, using INT8 precision during model conversion alongside a BF16 precision engine resulted in incorrect output tokens.

Moreover, to improve resource utilization and accelerate inference, engines are optimized for handling multiple requests under diverse concurrency configurations, which is also the major source of bugs, with 42 cases.
\emph{Concurrency Issue (RB.3)}, with 39 cases, implies the incorrect management of concurrent workers.
For instance, in Llama.cpp issue \#3537, the static variable `mu' used by the `Mirostat' sampler was shared across different workers, leading to state conflicts. 
This implies the need for engines to carefully address potential concurrency issues, including shared variables, atomic operations, and resource contention.
Moreover, \emph{Incorrect Job Context Management (RB.6)}, with 3 cases, refers to errors in handling context and configurations for multiple requests. 
For instance, in TensorRT-LLM issue \#1447, the engine fails to process requests with and without the `SamplingConfig' parameter simultaneously.

Moreover, \emph{Document Issue (RB.5)}, with 13 cases, refers to inaccurate or outdated descriptions of usage and configurations in documentation. For example, DeepSpeed issue \#4039 highlights the absence of inference configuration details for MoE models.

\finding{Managing configurations is challenging for LLM inference engines, accounting for 23\% of bugs across six root causes. Engines should prioritize comprehensive compatibility testing and timely updates for configuration combinations, model adaptation, mixed precision, concurrency conditions, multitasking, and documentation.}{finding:rc_configuration}

\subsubsection{[RC] Functionality}

Correct functionality implementation forms the operational backbone of LLM inference engines.
We identify 432 functionality-related cases (41 \% of total), spanning nine categories.
This substantial number highlights the challenges in balancing mathematical accuracy, framework compatibility, and performance optimization.
Next, we explain each root cause through examples.

Bug can originate from root causes that are common in DL libraries.
\emph{Incorrect Algorithm Implementation (RC.1)} refers to problematic functional implementation, such as the optimization strategies and computing kernels.
For instance, in vLLM issue \#590, an erroneous implementation of `RoPE' algorithm within attention operators caused unexpected output tokens.
This represents the most (43\%) prevalent functionality issue with 184 cases, emphasizing the complexity of functionalities in LLM inference engines.
\emph{Misused API (RC.2)}, with 75 cases, refers to incorrect usage of either internal or external APIs. 
This includes (1) failing to use an API or using it redundantly, and (2) using an incorrect API name or arguments.
For example, in vLLM issue \#9059, directly accessing a non-existent 'vocab' attribute in 'MistralTokenizer' instead of using 'get\_vocab()' caused a bug.
\revise{Additionally, RC.1 is used when the core problem lies in the algorithmic logic (e.g., wrong formula or algorithm adoption), while RC.2 is used when the bug stems from incorrect API usage (e.g., wrong arguments, wrong function calls). }
\emph{Mismatched Shape (RC.3)}, with 53 cases, occurs when tensor dimensions violate operator requirements.
For instance, in TensorRT-LLM issue \#2150, a `5D' tensor input to conv2d layers (which expect `3D/4D' inputs) led to a crash.

Bugs can stem from root causes that are widespread in software.
\emph{Incorrect Exception Handling (RC.4)}, with 37 cases, denotes failures to properly handle the exceptions.
For example, in Llama.cpp issue \#1768, the engine failed to raise exceptions for out-of-bounds input, resulting in a crash.
\emph{Incorrect Conditional Logic (RC.5)}, with 24 cases, involves incorrect handling of condition branches.
For instance, in Mlc-llm issue \#2361, the engine failed to handle the condition where `committed\_tokens' is empty during the dialog initialization phase, leading to a crash.
\emph{Incorrect Type (RC.6)}, with 24 cases, stems from type system violations.
For example, Deepspeed issue \#2183 mistakenly passed `ABCMeta' class objects instead of strings for name parameters.
\emph{Numerical Issue (RC.7)}, with 17 cases, involves failures in numerical computing, such as division by zero, overflow, and incorrect rounding.
For example, Deepspeed issue \#2090 treated microsecond timestamps as second units in flops profiling, causing metric inflation.
\emph{Incorrect Assignment (RC.8)}, with 15 cases, includes wrong variable/value assignments.
In Llama.cpp issue \#2050, a tensor dimension index was misassigned during OpenCL kernel configuration.

\emph{Incorrect Synchronization (RC.9)}, with 3 cases, is crucial to the efficiency of LLM inference engines, indicating parallel execution flaws.
For instance, vLLM issue \#6449 failed to synchronize `RNG' seed states across pipeline-parallel workers.

\finding{
The complex functionality of LLM inference engines contributes to 41\% of bug root causes.
The diversity of nine failure modes highlights the need for formal specification checking, comprehensive unit testing, and runtime validation across components of engines.
}{finding:rc_function}

\subsubsection{[RD] Environment}

Engines should adapt to various environments, which is a major source of bugs, with 301 cases (29\% of total), covering three leaf categories.

\emph{Incompatible Backend (RD.1)}, with 138 cases, indicates errors in implementations for specific backends.
For instance, in Llama.cpp issue \#1737, the memory copy operation of `ggml\_view\_3d' function encounters a crash due to the incompatibility of Metal backend.
The vast number of RD.1 highlights that the compatibility of various backends and devices is the major challenge for LLM inference engines implementations.
\emph{Incompatible Version (RD.2)}, with 98 cases, indicates incompatibilities arising from version changes in APIs. 
For example, in Mlc-llm issue \#2212, an incorrect TVM version led to an API call error because the TVM API had been updated.
This emphasizes the need for careful management of version and API changes in LLM inference engines.
\emph{Dependent Module Issue (RD.3)} involves errors or omissions in dependent libraries. 
For instance, in vLLM issue \#8107, the docker image of the engine lacked the `timm' dependency package. 
We observe 65 cases linked to this root cause, underscoring the challenge of managing extensive dependencies within the engine.

\finding{Environmental compatibility poses significant challenges for LLM inference engines, accounting for 29\% cases, spanning three categories.
This underscores the need for thorough compatibility testing, efficient version management, and reliable dependency management across various backends.}{finding:rc_environment}

\subsubsection{[RE] Resource}

LLM inference engines are generally optimized for efficient resource management, where we observe 37 cases (4\% of total) categorized into four leaf types.

\emph{Incorrect Resource Allocation (RE.1)}, with 14 cases, implies errors in allocating necessary resources. 
For example, in Llama.cpp issue \#52, a miscalculation in memory allocation led to an out-of-memory error despite available free memory. 
\emph{Incorrect Cache Management (RE.2)}, with 11 cases, involves flawed implementations in cache management. 
For instance, issue \#3825 in Llama.cpp demonstrates an incorrect shift operation within the KV cache.
\emph{Incorrect Multi-device Management (RE.3)}, with 8 cases, deals with errors in handling multiple devices.
In vLLM issue \#7472, the engine failed to detect varying CUDA compute capabilities across GPUs, causing misallocation in multi-GPU setups.
\emph{Incorrect Resource Deallocation (RE.4)}, with 4 cases, includes errors due to improper resource release. 
An example is TensorRT-llm issue \#1190, where resources were potentially released multiple times in IPC environments. 
RE.4 requires particular caution because it could lead to severe consequences like memory corruption or security vulnerabilities.

\finding{Resource management bugs, though only 4\%, are critical due to risks like memory leaks. 
We identify four leaf root causes, highlighting the need for carefully checking on resource allocation and release, cache management, and multi-device management.}{finding:rc_resource}

The substantial memory and computational requirements of LLMs lead to a pronounced prevalence of bugs stemming from two main root causes. 
The first category comprises bugs directly resulting from resource management mechanisms—such as KV-cache allocation or memory pool scheduling—which we classify as `Resource' (RE). 
The second category is `Functionality' (RC), which involves bugs where incorrect functional implementation induces resource-related problems, such as memory leaks due to flawed kernel logic. 
Empirical data suggests that bugs caused purely by resource management mechanisms are relatively infrequent, as RE accounts for only 4\% of observed cases. In contrast, the majority of resource-related bugs arise from faulty functional implementations, as functionality-related issues dominate our taxonomy, comprising 41\% of all bugs. 
These findings indicate that the notorious memory demands of LLMs significantly increase the complexity and vulnerability of functional implementations and resource management.

%

\begin{table}[t!]
\caption{\revise{Distribution of root cause types and the most frequent root causes across deployment phases. The numbers in parentheses indicate the number of issues and their respective percentages.}}
\label{tab:rc_dis}
\vspace{-1em}
\begin{center}
\resizebox{\linewidth}{!}{
\renewcommand{\arraystretch}{1.2}
\begin{tabular}{l r | >{\raggedright\arraybackslash}p{2.4cm} >{\raggedright\arraybackslash}p{2.4cm} >{\raggedright\arraybackslash}p{2.4cm} >{\raggedright\arraybackslash}p{2.4cm} >{\raggedright\arraybackslash}p{2.0cm}}
\toprule
\textbf{Phase} & \textbf{\# of Root} & 
\makecell[l]{\textbf{Top1}} & 
\makecell[l]{\textbf{Top2}} & 
\makecell[l]{\textbf{Top3}} & 
\makecell[l]{\textbf{Top4}} & 
\makecell[l]{\textbf{Top5}} \\
& \textbf{Cause Types} &&&&&\\
\midrule
\makecell[l]{\textbf{Engine} \\ \textbf{Setup}} & 12 & 
\makecell[l]{(RD.1) Incompa-\\tible Backend \\ (40, 26.0\%)} & 
\makecell[l]{(RD.3) Dependent \\ Module Issue \\ (35, 22.7\%)} & 
\makecell[l]{(RB.1) Misconfigu-\\ration \\ (30, 19.5\%)} & 
\makecell[l]{(RD.2) Incompa-\\tible Version \\ (22, 14.3\%)} & 
\makecell[l]{(RC.2) Misused \\ API \\ (6, 3.9\%)} \\
\midrule
\makecell[l]{\textbf{Model}\\ \textbf{Conversion}} & 15 & 
\makecell[l]{(RC.1) Incorrect \\ Algorithm \\ Implementation \\ (23, 14.6\%)} & 
\makecell[l]{(RB.2) Incompa-\\tible Model \\ (18, 11.4\%)} & 
\makecell[l]{(RB.1) Misconfigu-\\ration \\ (15, 9.5\%)} & 
\makecell[l]{(RD.1) Incompa-\\tible Backend \\ (13, 8.2\%)} & 
\makecell[l]{(RC.3) Mismat\\-ched Shape \\ (13, 8.2\%)} \\
\midrule
\makecell[l]{\textbf{Inference/}\\\textbf{Serving}} & 28 & 
\makecell[l]{(RC.1) Incorrect \\ Algorithm \\ Implementation \\ (158, 21.3\%)} & 
\makecell[l]{(RD.1) Incompa-\\tible Backend \\ (85, 11.5\%)} & 
\makecell[l]{(RD.2) Incompa-\\tible Version \\ (62, 8.4\%)} & 
\makecell[l]{(RB.1) Misconfigu-\\ration \\ (60, 8.1\%)} & 
\makecell[l]{(RC.2) Misused \\ API \\ (56, 7.6\%)} \\
\bottomrule
\end{tabular}
}
\end{center}
\vspace{-1em}
\end{table}

\revise{
\subsection{Root Cause Distribution}

Table \ref{tab:rc_dis} summarizes the distribution of root causes across different deployment phases.

Bugs in the \textbf{Engine Setup} phase stem from a total of 12 distinct types of root causes.
The primary root causes include \emph{(RD.1) Incompatible Backend}, \emph{(RD.3) Dependent Module Issue}, \emph{(RB.1) Misconfiguration}, \emph{(RD.2) Incompatible Version}, and \emph{(RC.2) Misused API}.
This indicates that the engine setup phase requires focused attention on ensuring environment compatibility and rigorous configuration management.

Bugs in the \textbf{Engine Setup} phase are caused by 15 unique root cause types.
The main root causes of bugs are \emph{(RC.1) Incorrect Algorithm Implementation}, \emph{(RB.2) Incompatible Model}, \emph{(RB.1) Misconfiguration}, \emph{(RD.1) Incompatible Backend}, and \emph{(RC.3) Mismatched Shape}.
This suggests that model conversion requires particularly careful algorithm implementation, configuration management, and environment adaptation.

Bugs in the \textbf{Inference/Serving} phase involve the highest number of root cause types, totaling 28. 
The predominant root causes are \emph{(RC.1) Incorrect Algorithm Implementation}, \emph{(RD.1) Incompatible Backend}, \emph{(RD.2) Incompatible Version}, \emph{(RB.1) Misconfiguration}, and \emph{(RC.2) Misused API}.
Similar to the model conversion phase, algorithm implementation is also a major challenge in the inference/serving phase, followed by environment adaptation and configuration management.

}

\subsection{Relationship Between Symptoms and Root Causes}

\begin{table}[h!]
\caption{Relationship between bug symptoms and root causes.}
\label{tab:relation}
\vspace{-1em}
\begin{center}
\resizebox{0.8\linewidth}{!}{
\begin{tabular}{l|rrrrrr}
\toprule
 & \textbf{Crash} & \textbf{Unexpected} & \textbf{Feature} & \textbf{Abnormal} & \textbf{System} & \textbf{Silent} \\
 & \textbf{ } & \textbf{Output} & \textbf{Failure} & \textbf{Perform.} & \textbf{Hang} & \textbf{Error} \\
\midrule
\textbf{Functionality} & 245 & 88 & 52 & 28 & 13 & 8 \\
\textbf{Environment} & 237 & 15 & 24 & 11 & 15 & 0 \\
\textbf{Configuration} & 153 & 26 & 33 & 14 & 14 & 4 \\
\textbf{In-/Output} & 16 & 13 & 9 & 1 & 2 & 0 \\
\textbf{Resource} & 22 & 3 & 3 & 7 & 1 & 1 \\
\bottomrule
\end{tabular}
}
\end{center}
\end{table}

To figure out the impact of each root cause and guide the location of root causes via symptoms, we explore the relationship between bug symptoms and root causes, as shown in Table~\ref{tab:relation}.

Considering the consequence of root causes, \emph{Crash} is the most common and harmful symptom across all five inner categories of root causes, making LLM services unavailable, highlighting the need for careful handling of all root causes.
Additionally, secondary symptoms vary across root cause categories.
Beyond crashes, bugs associated with \emph{Functionality} and \emph{In-/output} often lead to \emph{Incorrect Output}.
\emph{Environment} and \emph{Configuration} related issues are likely to cause \emph{Feature Failure} due to incompatibilities.
\emph{Resource}-related issues can cause \emph{Abnormal Performance}, because of the inefficient resource management.
These findings align with expectations and indicate the correctness of our taxonomies.

\finding{All root cause categories are significant due to their consistently high likelihood of causing crashes.}{finding:rc2sym}

For diagnosis of bug symptoms, first, the top three most common causes of \emph{Crash} are bugs related to \emph{Functionality}, \emph{Environment}, and \emph{Configuration}.
This indicates that when a crash occurs, developers should first check functionality errors and test for reproducibility across environments and configurations to quickly identify the cause.
Second, \emph{Feature Failure} and \emph{System Hang} are often equally linked to \emph{Environment}, \emph{Functionality}, and \emph{Configuration} issues, so developers can consider these categories when observing the two symptoms.
Third, for \emph{Incorrect Output} and \emph{Silent Error}, over half the bugs stem from \emph{Functionality} errors, highlighting the need to check functionality implementation for these symptoms.
Fourth, \emph{Abnormal performance} is mainly due to  \emph{Functionality} and \emph{Configuration} related issues.

\finding{
We identify following diagnostic patterns: crashes, feature failures, and system hangs require checks on functionality, environment, and configuration; incorrect output and silent errors need functionality checks; and abnormal performance involves functionality and configuration checks.
}{finding:sym2rc}

%% file: sections/commonality.tex

\section{RQ3: Commonality}
\revise{
In this section, we aim to explore the common characteristics of bugs across different engines, as well as the unique characteristics of bugs in each engine.
To achieve this, following prior work~\cite{islam2019comprehensive}, we compute the Pearson correlation coefficient ($r$) between engines. Specifically, we represent the distribution of bug symptoms and root causes in each engine as a normalized frequency vector, and then calculate the pairwise Pearson correlation between these vectors.}
Our primary hypothesis is that the engines will be strongly correlated based on the distribution of bug symptoms and root causes as they are doing similar tasks. 

\begin{figure}[h!]
    \centering

    \begin{subfigure}[t]{0.49\linewidth}
        \includegraphics[width=\linewidth]{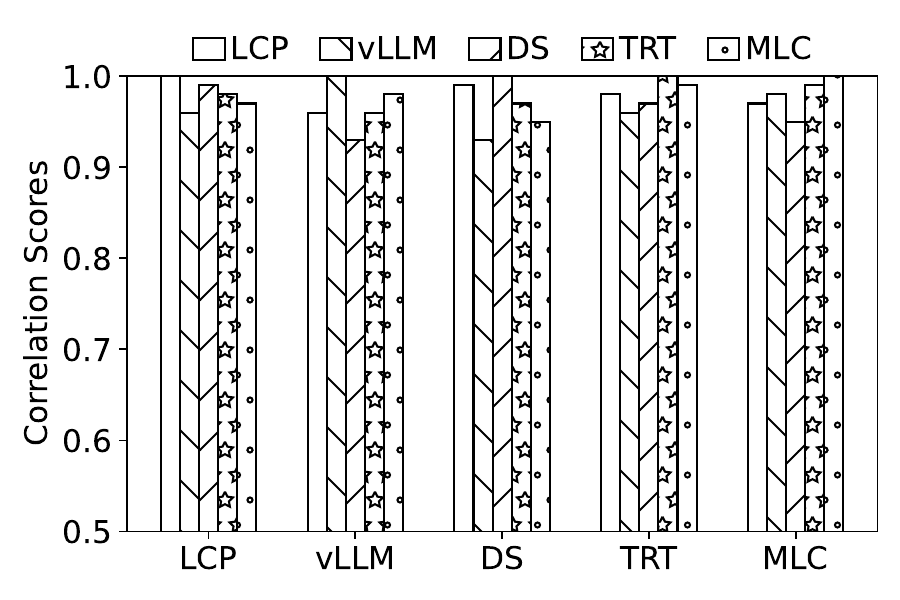}
        \caption{Symptoms.}
        \label{fig:correlation_symptom}
    \end{subfigure}
    \begin{subfigure}[t]{0.49\linewidth}
        \includegraphics[width=\linewidth]{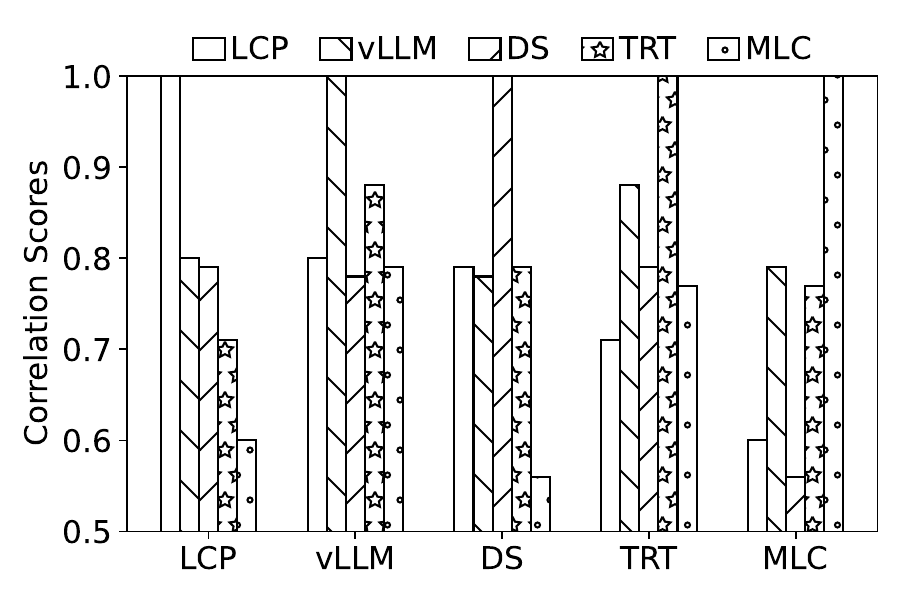}
        \caption{Root causes.}
        \label{fig:correlation_root_cause}
    \end{subfigure}

    \caption{\revise{Correlation of characteristics across engines. LCP = Llama.cpp, vLLM = vLLM, DS = DeepSpeed, TRT = TensorRT-llm, MLC = Mlc-llm. Correlations are calculated using the Pearson correlation coefficient over the distributions of symptoms (a) and root causes (b).}}
    \label{fig:correlation}
\end{figure}

Our analysis confirms that hypothesis.
\revise{As shown in Fig.~\ref{fig:correlation}, all the computed $r$ are greater than 0.5, indicating strong correlation~\cite{pearson} in terms of the distributions of both symptoms and root causes.}
These results indicate that the distribution of bug symptoms and root causes is consistent across different engines, highlighting the generalizability of our findings.
The correlation coefficients for bug symptoms are all above 0.9. 
This indicates that despite differences in targeted environments and implementations of LLM inference engines, bugs exhibit similar distributions and symptoms. 
This results further highlight that LLM app development requires a consistent bug tolerance method for LLM inference regardless of the engine used.

\finding{Bug symptoms and root causes across engines show a strong correlation ($r > 0.5$), indicating common bug patterns despite implementation differences.}{finding:commonality}

\begin{figure}[t!]
    \centering
    \includegraphics[width=0.9\linewidth]{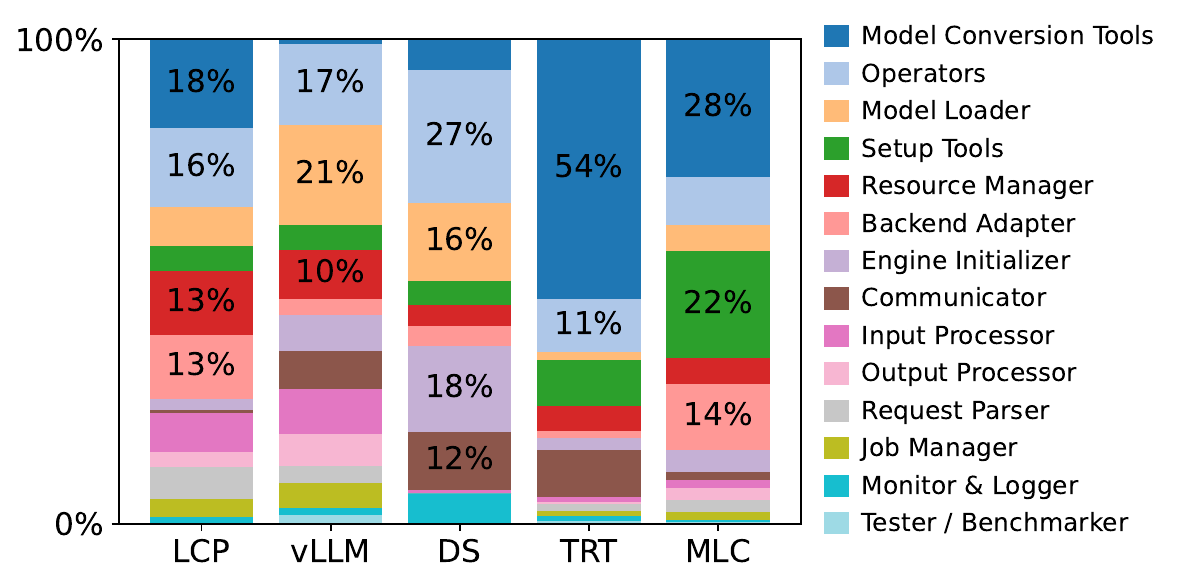}
    \caption{\revise{Prevalence of vulnerable components across engines. The y-axis represents the percentage of bugs attributed to each component. Proportions exceeding 10\% are annotated.}}
    \label{fig:components_engine}
    \vspace{-1.5em}
\end{figure}

Correlation coefficients for root causes are slightly lower than those for bug symptoms. 
This is likely due to differences in engine implementations.
Therefore, \revise{we further investigate the prevalence of vulnerable components across engines, i.e., the proportion of bugs attributed to each component, as shown in Fig.~\ref{fig:components_engine}, }
to explore potential implementation differences.

For engines designed for universal environments (including Llama.cpp and vLLM), the model conversion tool is the most bug-prone component. 
We observe a correlation coefficient of 0.64 between the vulnerable components of Llama.cpp and Mlc-llm, indicating their strong similarities.
Specifically, in Llama.cpp, the most vulnerable components are model conversion tools and operators, accounting for 18\% and 16\% of bugs. 
Model conversion tools are bug-prone due to the need of Llama.cpp to convert various original model formats (such as Huggingface model format) into a unified format for compatibility with diverse backends and devices. 
Operators of Llama.cpp are also bug-prone because they are written manually from scratch in C++ for each backend (e.g., CPU, Vulkan, and Metal), increasing the risk of bugs.
In Mlc-llm, model conversion tool issues account for 28\% of bugs due to its reliance on TVM for compiling models to specific backends. 
This compilation process involves complex optimizations, making it particularly bug-prone.

For engines tailored to cloud, like vLLM and DeepSpeed, operators and model loaders are the most bug-prone components. 
The correlation coefficient between vLLM and DeepSpeed is 0.66, also indicating strong correlation in their vulnerable components.
In vLLM, most bugs occur in the model loader and operators, accounting for 21\% and 17\%, respectively.
While vLLM supports seamless compatibility with Hugging Face model formats, it is vulnerable during model loading (such as distributed loading across devices) and with operators optimized for cloud.
Similarly, in DeepSpeed, bugs in operators and model loaders account for 27\% and 16\%, respectively.

TensorRT-LLM supports both cloud-based and edge server deployments.
During the model loading phase, beyond model format conversion and quantization, TensorRT-LLM also requires specifying parallel and KV cache optimization strategies, which are typically handled during inference/serving in other engines.
This complexity increases the bug rate for model conversion tools, accounting for 54\% of all TensorRT-llm bugs.

\finding{
The engines with the same targeted environment (e.g., universal or cloud) exhibit strong correlation ($r > 0.5$) in bug-prone components.
}{finding:component}

%% file: sections/repair_effort.tex
\revise{
\section{RQ4: Fix Effort}

To quantify the maintenance burden and guide prioritization of testing and development efforts, we investigate the fix effort required for different types of bugs in LLM inference engines.
Specifically, we analyze the fix duration for each bug issue in our dataset and examine how fix effort varies across bug symptoms and root causes, aiming to identify high-effort bug patterns that need focused attention.

To facilitate comparison, we classify the fix effort into four categories based on the fix duration: Easy ($\leq 1$ day), Medium ($(1, 7]$ days), Hard ($(7, 30]$ days), and Very Hard ($>30$ days). These correspond to fixes completed within one day, within one week, within one month, and exceeding one month, respectively.

\begin{figure}[h!]
    \centering

    \begin{subfigure}[t]{0.49\linewidth}
        \includegraphics[width=\linewidth]{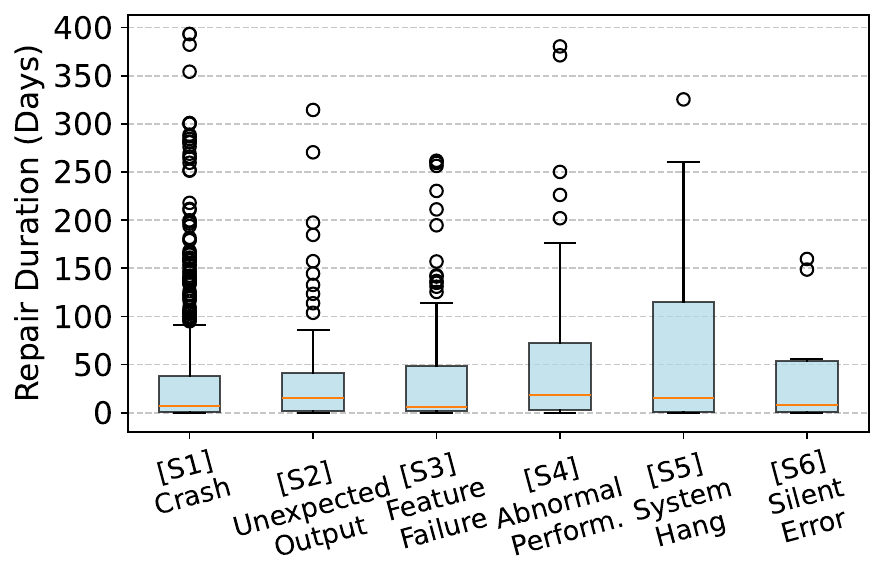}
        \vspace{-1em}
        \caption{Symptoms.}
        \label{fig:repair_symptom}
    \end{subfigure}
    \begin{subfigure}[t]{0.49\linewidth}
        \includegraphics[width=\linewidth]{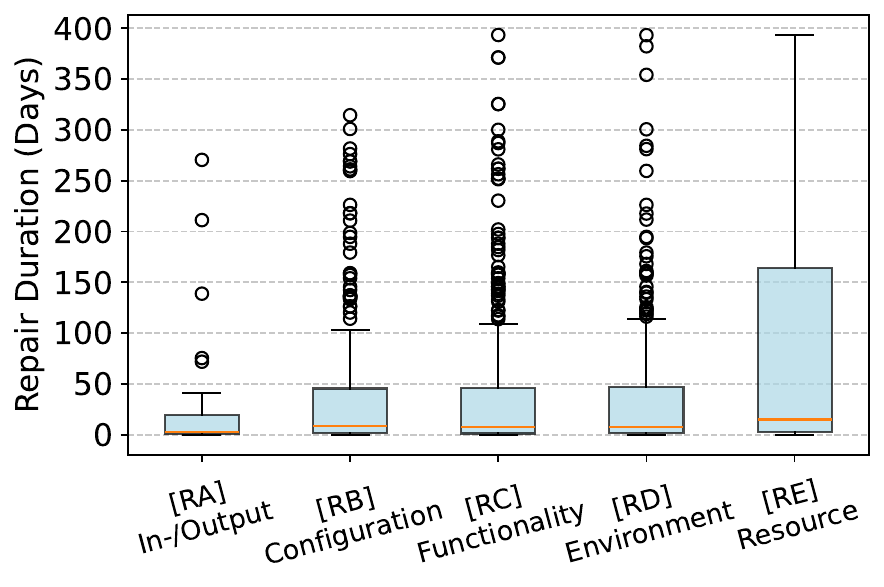}    \vspace{-1em}
        \caption{Root causes.}
        \label{fig:repair_root_cause}
    \end{subfigure}
    \vspace{-1em}
    \caption{Distribution of fix difficulty by symptoms (a) and root causes (b) of bugs.}
    \label{fig:repair}
\end{figure}

\subsection{Fix Effort by Symptoms}

Fig.~\ref{fig:repair}.(a) and Table~\ref{tab:symptom_repair_effort} show variations in the fix effort of bugs with different symptoms.
Crashes (S1), while prevalent, show a more balanced distribution across difficulty levels, with 28.1\% very hard cases and a median fix time of 7.2 days, though the high mean of 36.7 days fix duration indicates significant outliers. 
Unexpected output (S2) bugs exhibit a moderate challenge, with 34.4\% very hard cases and a median of 14.9 days, suggesting that issues like garbled tokens or semantic incoherence often require extended investigation.
Feature failures (S3) demonstrate a higher proportion of medium-difficulty cases (31.8\%), but still 28.0\% are very hard, with a low median of 6.4 days, hinting at quicker resolutions for some instances.
Abnormal performance (S4) stands out with the highest mean fix time of 58.2 days and 32.1\% very hard cases, reflecting the complexity of diagnosing performance degradation. 
System hangs (S5) are particularly demanding, with 42.9\% very hard cases and a high mean of 64.7 days, underscoring the difficulty of resolving unresponsive states. 
Silent errors (S6), though rare, show a balanced distribution, indicating variability in detection and fixes. 
Overall, non-crash symptoms like system hangs and abnormal performance consistently demand more effort, underscoring the need for diverse testing oracles beyond crash detection.

\finding{\revise{Non-crash bugs, such as system hangs, require substantial fix effort, highlighting the necessity of effective testing and fix strategies for non-crash bugs.}
}{finding:sym_repair}

\begin{table}[t!]
\centering
\caption{Distribution of fix difficulty by symptoms with percentage breakdown.}
\label{tab:symptom_repair_effort}
\vspace{-1em}
\resizebox{\linewidth}{!}{
\begin{tabular}{l|rrrr|rr}
\toprule
\textbf{Symptom} & \textbf{\# of Easy} & \textbf{\# of Medium} & \textbf{\# of Hard} & \textbf{\# of Very Hard} & \textbf{Mean (± Std)} & \textbf{Median} \\
 & \textbf{($\leq$1 day)} & \textbf{((1, 7] days)} & \textbf{((7, 30] days)} & \textbf{($>$30 days)} & (days) & (days)\\
\midrule
{[S1]} Crash & 133 (22.1\%) & 160 (26.6\%) & 140 (23.3\%) & 169 (28.1\%) & 36.7 (±66.2) & 7.2 \\
{[S2]} Unexpected Output & 31 (25.4\%) & 29 (23.8\%) & 20 (16.4\%) & 42 (34.4\%) & 33.6 (±51.8) & 14.9 \\
{[S3]} Feature Failure & 20 (18.7\%) & 34 (31.8\%) & 23 (21.5\%) & 30 (28.0\%) & 39.7 (±66.7) & 6.4 \\
{[S4]} Abnormal Perform. & 16 (30.2\%) & 13 (24.5\%) & 7 (13.2\%) & 17 (32.1\%) & 58.2 (±90.2) & 18.0 \\
{[S5]} System Hang & 4 (11.4\%) & 6 (17.1\%) & 10 (28.6\%) & 15 (42.9\%) & 64.7 (±91.4) & 15.7 \\
{[S6]} Silent Error & 2 (16.7\%) & 3 (25.0\%) & 3 (25.0\%) & 4 (33.3\%) & 38.0 (±57.8) & 8.2 \\
\bottomrule
\end{tabular}
}
\end{table}

\subsection{Fix Effort by Root Causes}

As shown in Fig.~\ref{fig:repair}.(b) and Table~\ref{tab:rc_repair_effort}, resource-related issues (RE) remain the most time-consuming, with 37.8\% very hard cases and a markedly high mean fix time of 79.5 days, coupled with a median of 15.0 days.
This underscores that the immense resource demands of LLM inference, such as significant memory requirements and multi-device coordination, significantly complicate the debugging process in LLM inference engines.
Configuration bugs (RB) and functionality issues (RC) show similar profiles, with 30.6\% very hard cases each and means of 42.3 and 39.6 days, respectively, indicating the huge overhead of managing complex settings and algorithm implementations. 
Environment-related causes (RD) follow closely, with 31.2\% very hard cases and a mean of 37.8 days, reflecting the effort needed for backend compatibility. 
In contrast, in-/output issues (RA) are relatively manageable, with only 19.5\% very hard cases and the lowest median of 2.5 days, suggesting that tokenization or formatting bugs can often be resolved swiftly.
These results underscores that resource and configuration root causes dominate the fix burden, necessitating prioritized testing and robust error handling in development workflows.

\finding{\revise{Resource and configuration issues are the primary fix burdens, necessitating prioritized testing and robust error handling in development workflows.}
}{finding:rc_repair}
}

\begin{table}[t!]
\centering
\caption{Distribution of fix difficulty by root causes with percentage breakdown.}
\label{tab:rc_repair_effort}
\vspace{-1em}
\resizebox{\linewidth}{!}{
\begin{tabular}{l|rrrr|rr}
\toprule
\textbf{Root Cause} & \textbf{\# of Easy} & \textbf{\# of Medium} & \textbf{\# of Hard} & \textbf{\# of Very Hard} & \textbf{Mean (± Std)} & \textbf{Median} \\
 & \textbf{($\leq$1 day)} & \textbf{((1, 7] days)} & \textbf{((7, 30] days)} & \textbf{($>$30 days)} & (days) & (days)\\
\midrule
{[RA]} In-/Output & 14 (34.1\%) & 11 (26.8\%) & 8 (19.5\%) & 8 (19.5\%) & 25.9 (±56.5) & 2.5 \\
{[RB]} Configuration & 46 (19.0\%) & 65 (26.9\%) & 57 (23.6\%) & 74 (30.6\%) & 42.3 (±70.2) & 8.8 \\
{[RC]} Functionality & 97 (22.5\%) & 113 (26.2\%) & 89 (20.6\%) & 132 (30.6\%) & 39.6 (±68.2) & 7.6 \\
{[RD]} Environment & 65 (21.6\%) & 82 (27.2\%) & 60 (19.9\%) & 94 (31.2\%) & 37.8 (±64.7) & 7.4 \\
{[RE]} Resource & 5 (13.5\%) & 9 (24.3\%) & 9 (24.3\%) & 14 (37.8\%) & 79.5 (±109.5) & 15.0 \\
\bottomrule
\end{tabular}
}
\end{table}

\subsection{\rerevise{Discussion on Fix Effort Metrics}}

\rerevise{In our study, we primarily use fix duration as a proxy for engineering effort. This metric captures the total turnaround time, which includes not only coding but also debugging, communication, testing, and review cycles. However, while this metric is commonly used in the software engineering community to measure fix efforts, we acknowledge that duration is an indirect measure, as developers' time may be split across tasks.

To provide a more comprehensive view, we conduct a complementary analysis using the number of changed lines of code (LoC), including added and deleted lines, from the fixing commits. This proxy measures the magnitude of the final code modification. As shown in \autoref{tab:loc_effort}, the LoC distribution is heavy-tailed (median is much smaller than the mean).

The results from the LoC analysis are consistent with our duration-based findings. Several non-crash symptoms show larger median code changes than crashes (e.g., `System Hang': 31 vs. `Crash': 17), confirming their complexity. For root causes, `Configuration' (RB) and `Resource' (RE) bugs exhibit a heavy tail, with extremely high means. This indicates that a notable subset of such bugs requires substantial refactoring, aligning with their potential for high fix duration.

Together, these two proxies provide a more robust understanding of fix effort. Fix duration captures the overall resolution overhead, while LoC highlights the scale of the implementation changes. These analyses reinforce the findings that non-crash symptoms and bugs rooted in resource or configuration issues represent a significant fix effort.}

\begin{table}[ht!]
\centering
\caption{\rerevise{Fix effort measured by changed lines of code (LoC).}}
\label{tab:loc_effort}
\resizebox{0.95\linewidth}{!}{
\begin{tabular}{l|rr|l|rr}
\toprule
\textbf{Symptom} & \textbf{Mean LoC} & \textbf{Median LoC} & \textbf{Root Cause} & \textbf{Mean LoC} & \textbf{Median LoC} \\
\midrule
{[S1]} Crash                & 6201  & 17   & {[RA]} In-/Output     & 83    & 30 \\
{[S2]} Unexpected Output    & 2812  & 32.5 & {[RB]} Configuration  & 16306 & 29 \\
{[S3]} Feature Failure      & 22563 & 29   & {[RC]} Functionality  & 7337  & 17 \\
{[S4]} Abnormal Perform.    & 59    & 17   & {[RD]} Environment    & 644   & 16 \\
{[S5]} System Hang          & 162   & 31   & {[RE]} Resource       & 2912  & 26 \\
{[S6]} Silent Error         & 98    & 25   &                       &       &    \\
\bottomrule
\end{tabular}
}
\end{table}

%% file: sections/fix_strategies.tex




\revise{
\section{RQ5: Fix Strategies}

We identify common fix strategies for bugs attributed to the five internal root cause categories.

\subsection{Fix Strategies for In-/Output Bugs}

\textit{Fix Input Validation.}  
This strategy, associated with 9 issues, addresses bugs caused by invalid, unexpected, or missing inputs through stricter input validation.
Solutions include validating request parameters and content (e.g., checking for empty or oversize inputs, skipping or sanitizing ill-formed bytes), enforcing required input formats (e.g., ensuring an ID field meets length/format constraints or rejecting unsupported extra fields), and providing graceful fallbacks or error messages. 
For instance, in Llama.cpp issue~\#7133, the fix replaces unsafe JSON subscript operators with safe ones and employs a robust assertion function for JSON parsing, which is not optimized out during compilation, to handle invalid JSON input.

\textit{Fix Input Format and Modality Support.}
This strategy, associated with 9 issues, enhances the system's ability to accept and accurately process new input types or formats, particularly for multi-modal content. 
Typical fixes include updating preprocessing logic to handle variable input shapes or sizes (e.g., resizing images or accommodating different patch counts for vision models), adjusting pipelines so that multi-modal data (like image or video inputs) are handled appropriately (e.g., preventing duplicate processing or overflow), and integrating previously unsupported models or features.
For example, in vLLM issue~\#5767, image inputs were uniformly resized to a fixed size to avoid errors caused by dynamic shape image inputs.

\textit{Fix Tokenization and Encoding.}  
This strategy, associated with 6 issues, involves refining text tokenization, detokenization, and character encoding. 
Typical solutions include aligning the tokenizer implementation with standard algorithms (e.g., ensuring consistent token breakdown), correcting how output text is pieced together (especially for multi-byte or special characters so they are not split or lost between tokens), skipping or adjusting calculations for tokens that could cause out-of-bound errors, and ensuring special tokens are recognized and processed properly (e.g., using the appropriate tokenizer files or combining vocab resources).
For instance, in Llama.cpp issue~\#167, the tokenizer was revised to correctly implement the SentencePiece tokenization algorithm, ensuring it generates standard-compliant token sequences.

\subsection{Fix Strategies for Configuration Bugs}

\textit{Fix Model and Feature Support.} 
This strategy, associated with 29 issues, focuses on extending or modifying the system to accommodate new models, architectures, or functionalities that were previously unsupported. Solutions include adding compatibility for additional model types or formats (e.g., integrating new model architectures or supporting new quantization schemes), implementing missing operations or components needed by certain models (e.g., adding a CUDA kernel for a specific activation function or enabling LoRA or adapter support for a model), and adjusting the code to handle unique model configurations or hardware scenarios (e.g., updating GPU-specific logic so that older GPUs or CPU-only deployments are properly supported). 
For instance, in Llama.cpp issue \#4038, adding missing CUDA kernels for ReLU and SQR operations allowed the Persimmon model to successfully offload computations to the GPU, resolving the crashes that occurred previously due to those unsupported operations.

\textit{Fix Configuration Defaults.}  
This strategy, associated with 13 issues, improves default settings and configuration management to prevent misconfigurations and ensure stable operation. Solutions include introducing new configuration options or environment variables (e.g., adding a `MAX\_JOBS' limit for build processes or a setting to cap concurrent requests), and adjusting default values to safer or more optimal levels (e.g., reserving extra GPU memory by default to prevent out-of-memory errors). 
For instance, in vLLM issue \#9797, the `host' argument was fixed to default to 0.0.0.0 instead of the previous default value of `None', preventing the service from listening on an unintended network interface.

\textit{Fix Backward Configuration Compatibility.} 
This strategy, associated with 7 issues, ensures that updates do not break existing workflows by maintaining compatibility with legacy configurations, formats, or behaviors. 
Solutions include supporting legacy file formats or parameters alongside new ones (e.g., accepting older space-delimited tokenizer merge files in addition to the new list format), preserving functionality with older library versions or usage patterns (e.g., adapting to both Pydantic v1 and v2 syntax changes), and handling extra or missing data more leniently to maintain previous behavior (e.g., skipping over non-critical tensors in a newer model checkpoint or turning a potential error into a warning so that older model files still load). 
For instance, in Llama.cpp issue~\#9692, the engine was updated to accept both the new version pair-list and old version space-delimited tokenizer configuration files.

\textit{Fix Documentation.} 
This strategy, associated with 9 issues, involves updating and improving documentation and examples to guide users correctly and prevent confusion-related errors. 
Solutions include adding or expanding explanations for complex features (e.g., documenting how to configure asynchronous engine parameters or multi-node inference setups), updating any outdated or incorrect information (e.g., fixing broken download links or correcting deprecated parameter names in guides), and providing clearer examples for common pitfalls (e.g., demonstrating how to use a model’s full context length or reminding users to unset debug environment variables after troubleshooting). 
For instance, in vLLM issue \#7815, the documentation was updated to clearly explain how to set the `VLLM\_HOST\_IP' environment variable for multi-node inference, helping users avoid misconfigurations in distributed setups.

\subsection{Fix Strategies for Functionality Bugs}

\textit{Fix Algorithm Logic.} 
This strategy, associated with 66 issues, addresses functional bugs or inefficiencies caused by flawed algorithm implementations or incorrect mathematical logic by correcting and refining the core procedure. 
Typical solutions include adjusting key computational steps to ensure correct outcomes, e.g., fixing the order of operations in inference or search algorithms, adding missing normalization or transformation steps, or properly handling random seed initialization to achieve consistent behavior. 
For example, in Llama.cpp issue \#186, the computation algorithm for the `RMS' norm operator was introduced.

\textit{Fix API Usage and Compatibility.} 
This strategy, associated with 15 issues, fixes bugs arising from improper use of external APIs or interfaces (e.g., deprecated function calls, missing calls, or incompatibilities with newer library versions) by updating the code to align with the expected API usage. 
Typical solutions include replacing outdated or deprecated function calls with their updated counterparts, adding required parameters or method calls to meet new interface requirements (e.g., calling a flush function after writing to a log to force immediate output), and modifying code to adapt to changes in third-party libraries. 
For example, in Llama.cpp issue \#3753, a segmentation fault was resolved by replacing outdated sequence APIs with newer batch APIs to match the updated library design.


\textit{Fix Data Type Handling.}
This strategy, associated with 9 issues, addresses bugs caused by improper data types or precision issues by ensuring code variables use appropriate and consistent types. 
Typical fixes include using larger or more suitable data types to prevent overflow (e.g., changing a byte-counter from a 32-bit integer to a 64-bit long so it can handle very large values) and unifying the data type of tensors or variables across different parts of a system to avoid type-mismatch errors. 
For example, in Llama.cpp issue \#6654, an integer overflow occurred because an `int' was used to calculate file byte sizes for a model-splitting feature. The fix changed this counter to a `long' type, which can represent much larger numbers, thereby preventing the overflow for large model files.

\textit{Fix Tensor Shape Alignment.}
This strategy, associated with 10 issues, addresses bugs arising from mismatched tensor dimensions or shapes.
Typical solutions include validating tensor dimensions before operations and adjusting them if necessary(e.g., adding checks to ensure a matrix has the required minimum number of columns for multiplication), padding tensors with zeros to meet alignment requirements (e.g., extending length to the next power of two to satisfy a GPU kernel’s constraints), or reorganizing how data is partitioned so that each compute device (GPU/TPU) gets correctly shaped slices of the overall tensor. 
For instance, in vLLM issue \#4127, a GPU attention kernel required the head dimension to be a power of two.
The fix zero-padded the tensors up to the next power-of-two length, ensuring the kernel could run without errors regardless of the original tensor shape.

\textit{Fix Control Flow.}
This strategy, associated with 24 issues, fixes incorrect or unstable program behavior by adjusting the code’s branching and execution order, adding necessary conditions, and avoiding invalid execution paths. 
Typical solutions include inserting missing conditional checks or guard clauses to handle special cases (e.g., skipping a block of code when a required flag or token ID is None to prevent running invalid logic), reordering operations to follow a safe sequence, and introducing asynchronous or conditional execution where appropriate. 
For instance, in Llama.cpp issue \#3766, synchronous updates to a system prompt were blocking server response handling. The fix changed the control flow to update the system prompt asynchronously, decoupling it from the main request processing. This change allowed the server to handle concurrent requests smoothly, eliminating delays and deadlocks.

\subsection{Fix Strategies for Environment Bugs}

\textit{Fix Dependency Management and Version Alignment.}
This strategy, associated with 23 issues, updates or aligns software dependency versions. 
Solutions include bumping library versions to incorporate upstream fixes, relaxing or adjusting version constraints, or rolling back problematic updates to resolve conflicts. 
For instance, in vLLM issue \#4509, the fix upgraded the project’s `PyTorch' dependency to the newest version, which eliminated a version conflict with the `Triton' library. 

\textit{Fix Build and CI Pipeline.}
This strategy, associated with 9 issues, focuses on refining build scripts, CI configurations, and packaging processes to ensure software can be compiled and deployed reliably. 
Typical solutions include adjusting continuous integration (CI) workflows or build scripts (e.g. Dockerfiles, CMakeLists, GitHub Actions YAML) to correct file paths, environment variables, or tool versions, and reorganizing artifact outputs so that releases are packaged correctly (seen in Llama.cpp issue \#6604). 

\textit{Fix Cross-Platform Compatibility.}
This strategy, associated with 22 issues, adapts the codebase to run consistently across different operating systems, hardware architectures, or runtime environments by addressing platform-specific issues. 
Solutions include adding conditional logic or using portable APIs to handle OS-specific differences, adjusting compiler directives or flags (seen in Llama.cpp issue \#2922), and extending support for various hardware (like different GPU architectures or CPU instruction sets) and language runtimes. 

\textit{Fix Missing Dependencies and Components.}
This strategy, associated with 6 issues, ensures all required dependencies and components are present so the software doesn’t fail due to missing modules or files. 
Common fixes include adding omitted libraries or packages to installation requirements (or Docker images), restoring files accidentally left out of distributions, and adjusting setup scripts to include all necessary components. 
For example, in Mlc-llm issue \#1547, an absent Python submodule register file `\_\_init\_\_.py' was added, preventing module-not-found errors in new environments.

\textit{Fix Documentation for Environment Setup.}
This strategy, associated with 7 issues, improves installation and configuration documentation to help users set up their environment correctly and avoid known pitfalls. 
Solutions include clarifying platform-specific setup requirements (seen in Mlc-llm issue \#421), providing step-by-step setup instructions, adding notes about required tools or versions, and compiling FAQs or troubleshooting tips for common errors. 

\subsection{Fix Strategies for Resource Bugs}

\textit{Fix Memory Management.}
This strategy, associated with 7 issues, addresses memory safety and efficiency issues.
Typical solutions include adding boundary checks for array and buffer accesses (especially in low-level GPU kernels, seen in vLLM issue \#4756), freeing temporary buffers immediately after use, caching and reusing expensive results to prevent redundant memory allocations (e.g., avoiding repeated initialization of large GPU contexts), and adjusting feature toggles or scheduling to keep memory usage within limits. 
}

%% file: sections/temporal.tex
\revise{
\section{RQ6: Temporal Evolution}

In this section, we investigate the temporal evolution of bug types over time.
For each issue, we use the creation time as the reference timestamp. 
To mitigate the influence of varying development speeds and issue reporting rates, we avoid partitioning the timeline using fixed calendar intervals. 
Instead, we define each period bin to contain exactly 50 consecutive bug reports, ordered by creation time. This issue-count-based binning normalizes the analysis across fluctuating development intensities and ensures each period reflects a comparable volume of engineering activity. 
To examine sustained trends, we focus on the vLLM engine, which offers the largest sample size of 290 issues, yielding seven consecutive periods from May 2023 to November 2024.

\begin{figure}[t!]
    \centering
    \includegraphics[width=0.8\linewidth]{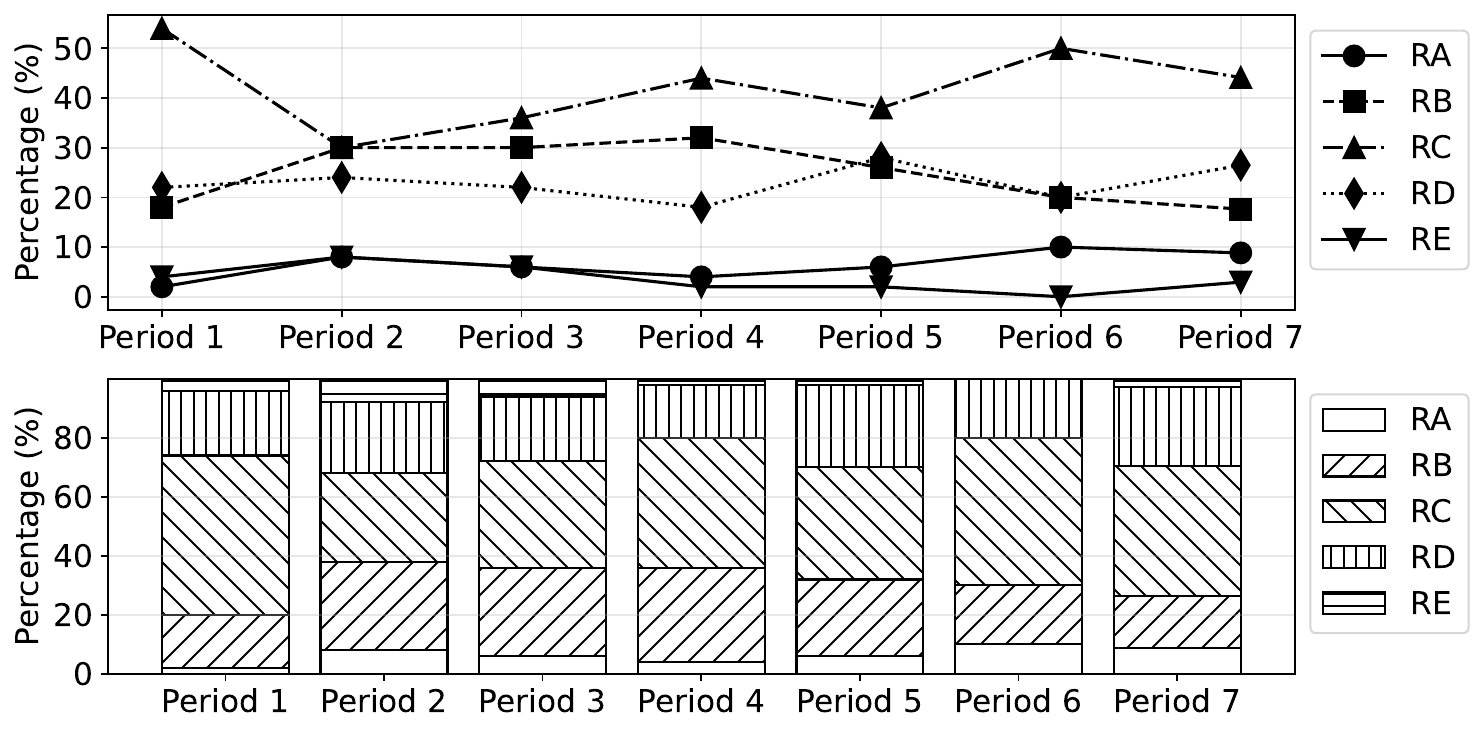}
    \caption{Temporal evolution of root cause categories percentage distribution in vLLM. Each period contains 50 consecutive bug issues, ordered by their creation time.}
    \label{fig:temporal}
    \vspace{-1.5em}
\end{figure}


As shown in Fig.~\ref{fig:temporal}, during the early stages of development (Period 1), functionality-related (RC) issues constitute the largest proportion of reported bugs. This trend aligns with typical software development patterns, where initial efforts focus on implementing core functionality. Although the proportion of RC-type bugs gradually decreases in subsequent periods, they remain the dominant category throughout the observed timeline.
Configuration-related (RB) bugs experience an increase during the mid-phase of the observed timeline (Periods 2--4), coinciding with the expansion of configurable options and features in the inference engine. As the set of configuration options stabilized in the later periods, the proportion of RB-type bugs decreased.
Environment-related (RD) issues maintained a relatively stable presence across all periods. 
In-/Output-related (RA) and resource-related (RE) bugs, although consistently comprising a smaller fraction of the total, intermittently emerge and resist complete resolution.

}

%% file: sections/discussion.tex
\section{Discussion}
\label{sec:discussion}

\subsection{Comparison with Prior Work on DL System Bugs} 

To elaborate on the comparison with existing literature, we analyze the differences between our findings and findings on DL system bugs in prior work~\cite{zhang2018empirical,humbatova2020taxonomy,zhang2020empirical,islam2019comprehensive,chen2023toward}.
\revise{While LLM inference shares certain similarities with traditional deep learning (DL) inference, allowing for the occurrence of some DL-related or general software bugs, it also possesses unique characteristics.
These stem from the large model scale and specialized optimization techniques inherent to LLMs. Additionally, LLM inference engines are specifically designed to focus on inference, unlike earlier DL libraries that often support both training and inference. 
These distinct features contribute to the unique nature of bugs found in LLM inference engines.}

First, LLM inference shows unique bug symptoms and distribution.
(1) Unexpected output is the second most frequent and unique bug symptom (Finding~\ref{finding:symptoms}) rarely observed in prior DL inference studies~\cite{zhang2018empirical,humbatova2020taxonomy,zhang2020empirical,islam2019comprehensive,chen2023toward}. This is because LLMs are expected to generate human-readable output, making unreadable token bugs more noticeable and common. In contrast, traditional DL models, like CNN classifiers, produce single classification results, where errors are often attributed to model capability instead of bugs; therefore these bugs are rare and cannot form a distinct symptom category. 
(2) LLM inference does not show bug symptoms related to data preparation and training, which are dominant in traditional DL studies~\cite{humbatova2020taxonomy,islam2019comprehensive}. This difference arises because traditional DL systems handle both training and inference together, whereas LLM inference engines focus solely on inference and serving, leading to distinct bug symptoms and distributions.

\revise{Second, LLM inference shows unique bug root causes and distribution.
(1) 7 identified bug root cause categories are not prominent enough to form distinct bug categories in prior DL studies~\cite{humbatova2020taxonomy,islam2019comprehensive}, including incorrect request parsing, insufficient multimodal support, incorrect formatting, incorrect job context management, incorrect synchronization, incorrect cache management, and incorrect resource deallocation. 
These root causes stem from LLMs' unique requirements that are absent in general DL libraries, including multi-tenant serving, multimodal processing, output format control, a vast space of optimization configurations, distributed resource management, and KV cache optimization needs.
(2) Shared bug root causes have a unique distribution in LLM inference. For example, environment-related root causes account for 29\% in LLM inference, compared to just 8\% in general DL libraries~\cite{chen2023toward}. This is because LLM inference engines typically support multiple deployment environments within a single engine, whereas DL systems usually target a single environment, such as TF-serving for the cloud, TF Lite for mobile, and TF.js for browsers~\cite{chen2020comprehensive}.}

\rerevise{Additionally, recent work studies faults in attention-based neural networks and proposes attention-specific fault taxonomies that cover both training and inference stages~\cite{jahan2025attention}. Their focus lies on the model/architecture layer to explore how attention mechanisms fail across projects and frameworks. In contrast, our work focuses on the engine layer, which treats LLM inference engines as deployment and serving systems. We analyze bugs across deployment phases (setup, model conversion, inference/serving) and characterize system-oriented causes such as environment/backend incompatibility, configuration interactions, request/job management, and resource scheduling. Therefore, the two lines of work are complementary. Attention-specific taxonomies explain how attention modules fail, while our taxonomy explains how inference engines fail when deploying and serving LLMs under diverse environments and configurations.}

\subsection{Implications}
With the rising popularity of LLM inference engines, our study offers timely and practical insights for different stakeholders.

\noindent\paraf{For inference engine vendors}, our taxonomy in Finding~\ref{finding:rc_taxonomy} outlines 28 unique root causes, helping vendors understand and quickly address or avoid these issues. 
Findings~\ref{finding:rc_inoutput}--\ref{finding:rc_resource} highlight areas that require special attention during implementations. 
For example, Finding \ref{finding:rc_inoutput} advises developers to concentrate on checking, processing, and error handling for character encoding, tokenizers, and user-specific formatting when dealing with inputs and outputs.
While vendors might consider prioritizing different root causes, Finding~\ref{finding:rc2sym} advises treating all bugs with equal seriousness, since any could pose a high risk of causing crashes.
Additionally, developers might wonder which engine components are more prone to errors.
This question is answered by Finding \ref{finding:component}.
For example, for engines targeting cloud environments, special focus should be on the implementation of model loaders and operators.

\noindent\paraf{For LLM app developers}, Finding \ref{finding:symptoms} highlights the symptoms to consider for fault tolerance during the development of LLM apps, including crashes, unexpected outputs, feature failures, abnormal performance, system hangs, and silent errors.
Furthermore, Findings \ref{finding:sym_es}--\ref{finding:sym_phases} recommend stage-specific fault tolerance strategies for LLM app developers. For example, Finding \ref{finding:sym_es} reveals that 97\% of bugs appear as crashes during the setup phase, highlighting the importance of implementing crash fault tolerance during LLM apps setup.
Moreover, Finding \ref{finding:commonality} shows that symptom distribution is consistent across engines, indicating that fault tolerance considerations are the same regardless of the engine used in LLM apps.
Additionally, Finding \ref{finding:rc_environment} identifies environmental compatibility as a major (29\%) error source, urging LLM app developers to clearly specify target environments and select suitable engines.

\rerevise{For debugging memory-related bugs, our study provides a key diagnostic heuristic: Memory problems (e.g., memory leaks, OOM, abnormal memory growth) may not solely originate from bugs in resource management modules. Rather, due to the immense resource demands of LLM inference, incorrect functional logic can lead to excessive resource allocation during implementation or a failure to deallocate resources after an operation is complete. 
For developers, this suggests a strategy of investigating both the logic layer and the resource management layer in parallel when encountering memory-related bugs. While inspecting the memory allocator or KV cache scheduler, developers should place a strong emphasis on investigating the functional implementation, such as custom CUDA kernels, complex attention mechanisms, or sampling algorithms. For instance, a bug is more likely to be an incorrect tensor shape calculation (RC.3 Mismatched Shape) causing an out-of-bounds write, or a flawed algorithm (RC.1 Incorrect Algorithm Implementation) that fails to release memory on an edge case, than a bug in the memory pool manager itself.}


\noindent \paraf{For researchers}, our study provides a foundational empirical basis that opens several specific and actionable research directions. 
\rerevise{Finding \ref{finding:token_oracle} is the first to empirically identify the features of bugs under LLM undetermined output tokens, providing concrete insights for researchers to design test oracles that move beyond simple crash detection. For instance, these factors can be translated into concrete automated oracles. 
\textbf{Inconsistent outputs} can be detected via a reproducibility oracle that runs the same prompt multiple times under deterministic decoding (e.g., greedy decoding with fixed seeds) and asserts identical outputs.
\textbf{Abnormal length} and \textbf{repetition} can be flagged by simple checks, such as empty or single-token outputs, max-token exhaustion, high repeated $n$-gram rates (e.g., $n{=}3$--$5$), or consecutive substring repetition.
\textbf{Mis-encoded characters} can be identified by validating that outputs are well-formed Unicode/UTF-8 and by searching for the Unicode replacement character (e.g., U+FFFD) or unexpected control characters.
For judging \textbf{incoherent semantics}, more advanced methods can be explored, such as employing a powerful external LLM as a judge to evaluate the semantic coherence of the generated text. 
These examples demonstrate how our findings enable the development of practical, automated techniques to validate LLM outputs at scale.}
Additionally, Finding \ref{finding:sym_distribution} indicates that researchers should not rely solely on crashes as the oracle for LLM engine testing, as over 35\% of bugs exhibit other symptoms, underscoring the need for a diverse set of oracles that encompass performance anomalies, feature failures, and silent errors to achieve comprehensive testing coverage. 

Furthermore, Findings \ref{finding:rc_inoutput}--\ref{finding:rc_resource} offer a robust foundation for developing more efficient and targeted testing methods by highlighting key vulnerability areas. 
For instance, Finding \ref{finding:rc_configuration} reveals that many bugs are triggered under complex configuration conditions, such as model compatibility, mixed precision, and concurrency, highlighting the critical need for configuration-aware testing. Researchers could design advanced configuration-space exploration techniques specifically for LLM inference engines, such as combinatorial interaction testing or adaptive sampling, to efficiently navigate the high-dimensional parameter space of LLM engines, thereby improving test efficiency by focusing on high-risk combinations that are prone to errors. 
Similarly, Findings \ref{finding:rc_function} and \ref{finding:rc_environment} point to the critical need for testing environmental compatibility and resource management, advocating for techniques like static analysis for resource leak detection and metamorphic testing across backends. 
By addressing these engine-specific challenges, researchers can advance the state of testing for LLM inference systems, ensuring their reliability and performance across diverse deployment scenarios.

Moreover, the strong commonality of bug patterns across different engines, as evidenced by Finding \ref{finding:commonality}, indicates a significant level of cross-engine generalizability. 
This suggests that researchers may investigate whether oracle definitions and testing techniques, such as those based on output validation or configuration testing, can be generalized across inference engines, enabling the creation of reusable research artifacts. 
For example, standardized test suites or metamorphic relations derived from one engine could be adapted to others, reducing duplication of effort and accelerating the validation of new or updated engines. By leveraging these shared patterns, researchers can develop more scalable and portable methods for ensuring the reliability of LLM inference systems in diverse environments.

\subsection{General Guidelines for Developing LLM Inference Engines}

Based on our findings, we aim to provide general guidelines for developing LLM inference engines structured by the composited modules.

\revise{
\noindent\paraf{Engine Setup Module.} 
The engine setup phase is critical for ensuring a stable foundation, as bugs at this stage often prevent deployment entirely.
(1) LLM engine developers can conduct thorough cross-platform testing across all supported backends, platforms, and devices, as Finding \ref{finding:rc_environment} implies that 29\% of issues arise from these factors. 
For instance, continuous integration pipelines should include targeted testing on heterogeneous CUDA environments (vLLM issue\#7472) and the Metal backend of Apple Silicon (Llama.cpp issue \#1737), validating core functions like tensor operations and memory management.
(2) Automated dependency checkers can be employed to ensure compatibility of dependencies and installation configurations, as Finding \ref{finding:rc_configuration} indicates that incorrect setup configuration is a major root cause.
For example, these checkers should be designed to validate critical configuration aspects, such as ensuring consistency in parallelism configurations across distributed settings (DeepSpeed issue \#3628), maintaining precision alignment between models, engines, and inputs (TensorRT-LLM issue \#1303), verifying the completeness of dependencies for specific deployment modes (vLLM issue \#8107), and checking the availability of system libraries and headers during source compilation (Llama.cpp issue \#279).
(3) Developers can prioritize detecting crash symptoms because Finding \ref{finding:sym_es} suggests most bugs in this phase lead to crashes.

\noindent\paraf{Model Conversion Module.}
(1) LLM engine developers can implement validation steps during and after conversion (e.g., checksums and precision checks) to ensure converted models are complete and configured correctly, as Finding \ref{finding:sym_mc} suggests that generated model files can be flawed. 
For example, validators should check for complete parameter sets and correct precision after AWQ quantization (TensorRT-LLM issue \#1245) to avoid silent errors during inference.
Moreover, developers can establish a comprehensive testing matrix covering critical configuration combinations. This matrix should systematically evaluate interactions between model architectures and formats(e.g., TensorRT-LLM issue \#1634), quantization methods and precision settings (e.g., testing INT4, AWQ, and GPTQ across different precision backends), and target backend platforms (e.g., validating CUDA, Metal, and Vulkan conversions for each model type).
Such combinatorial testing can proactively identify incompatibilities that lead to crashes, incomplete outputs, or misconfigured models.
(2) Given that Finding \ref{finding:component} indicates model conversion tools are highly bug-prone, developers can regularly update and test these tools for their compatibility, correctness, and performance.
For instance, test suites should cover edge cases like tensor shape mismatches during SmoothQuant conversion (TensorRT-LLM issue \#899) and validate output model compatibility across backends.

\noindent\paraf{Inference/Serving Module.}
(1) LLM engine developers can implement multiple testing oracles, as suggested by Findings \ref{finding:sym_phases} and \ref{finding:sym_distribution}. 
For example, validation of output tokens can be incorporated to check features described in Finding \ref{finding:token_oracle} (e.g., character correctness, semantic coherence, repetition). 
(2) Because Finding \ref{finding:symptoms} implies that abnormal performance is a major bug symptom in this phase, developers can conduct performance tests under different resource settings.
For example, performance tests should measure throughput and latency under heterogeneous LLMs (DeepSpeed issue \#2488) and validate KV cache management under memory constraints.
(3) Building on Finding \ref{finding:rc_configuration}, which reveals that configuration-related bugs comprise 23\% of all issues, it is essential to formulate a rigorous testing strategy that specifically targets high-risk configuration combinations in the inference/serving stages. 
This strategy should center on designing comprehensive testing matrices to evaluate critical interactions across multiple dimensions, including concurrency settings, resource limits, job parameters, model types, backend devices, and parallel configurations.
For example, a focused approach involves testing various worker counts under constrained KV cache sizes to preempt shared state conflicts with concurrent workers (Llama.cpp issue \#3537). 
This should be extended to experimenting with diverse job parameters, such as max\_tokens values, sampling options, and grammar constraints applied across a range of model architectures. 
Additionally, validation must ensure compatibility in multi-GPU environments with tensor parallelism enabled (vLLM issue \#7472), and verify the correct synchronization of parallel workers to prevent performance anomalies. 
}



\subsection{Threats to Validity}
First, our analysis focuses on five LLM inference engines, which may limit the generalizability of our findings to all LLM inference engines.
To mitigate this threat, we carefully selected engines with high popularity, well-known providers, and diverse targeted environments.
Second, our findings are based on reported issues from GitHub, potentially introducing bias. 
To mitigate this, we collected a large number of issues, specifically 929 bug-related issues, through careful manual inspection. 
\revise{Third, our methodology involves manual bug categorization and the two inspectors involve in the coding process are doctoral students, which may introduce bias. 
To mitigate this, three authors with at least over two years of LLM apps development participate in taxonomy construction, resolving inconsistencies through discussion until reaching an agreement.
Moreover, the final Kappa values for the symptom and root cause labeling are greater than 0.8, indicating almost perfect agreement and confirming the reliability of the coding schema.}
\rerevise{Fourth, this study focuses only on closed issues to ensure sufficient evidence for reliable characteristic labeling and analysis. The trade-off of this methodological choice is that we may omit newly reported or unresolved problems in the open issues, which might include recent challenges as LLM inference engines evolve rapidly. 
Future work could complement this study by incorporating open issues once they are resolved or better documented.}


%% file: sections/relatedwork.tex
\section{Related Work}
\label{sec:relatedwork}

\paraf{Empirical study on bugs in machine learning libraries.}
Bugs in machine learning systems have long been a significant concern in software engineering~\cite{guo2019empirical,chen2020comprehensive,zhang2018empirical,chen2023toward,guan2023comprehensive,islam2019comprehensive}. 
Chen et al.~\cite{chen2020comprehensive} have conducted a comprehensive survey and analysis on Stack Overflow of developers' questions about deploying machine learning models in the cloud, mobile, and browsers, establishing a taxonomy of specific deployment challenges. 
Chen et al.~\cite{chen2023toward} have conducted an empirical study on four popular deep learning frameworks and measured the test coverage achieved by three testing techniques. 
While previous studies have investigated bugs in machine learning deployments across various platforms, they primarily focus on traditional machine learning models. 
Our work focuses on bugs in LLM inference engines, which have unique characteristics, such as undetermined output, high resource demands, and compatibility for diverse environments.

\paraf{Testing of machine learning libraries.}
Many studies have proposed different methods to test machine learning frameworks, including compilers~\cite{liu2023nnsmith, liu2022coverage, ma2023fuzzing,luo2021graph} and libraries~\cite{gu2022muffin, luo2021graph, wang2020deep, li2023comet, pham2019cradle, yang2023fuzzing, wei2022free, deng2023large, deng2024large}. 
Liu et al.~\cite{liu2023nnsmith} have proposed NNSmith, which leveraged symbolic constraint solving and gradient-based search to generate diverse yet valid DNN test models for fuzzing deep learning compilers. 
Additionally, some studies leverage large language models for testing deep learning libraries~\cite{deng2023large,deng2024large}. 
TitanFuzz~\cite{deng2023large} leverages Large Language Models (LLMs) for generating valid input programs to fuzz deep learning libraries. 
Although existing work has explored various components of DL frameworks, no studies have specifically addressed the bugs in LLM inference engines.
Our work provides an empirical foundation for future research on LLM inference engines.
Moreover, our open-source dataset can serve as a data source for enhancing the LLM-based testing methods.

%% file: sections/conclusion.tex
\section{Conclusion}
\label{sec:conclusion}

\revise{
We present the first study on bugs in LLM inference engines, addressing a critical gap in understanding the bug symptoms, root cause, commonality, fix effort, fix strategies, and temporal evolution of bugs.
}
By examining five well-known engines, we identify six bug symptoms, 28 root causes, and strong correlation of these characteristics across engines, offering insights into efficient bug detection and location.
Our primary findings include that (1) Five factors are useful for oracle setting under undetermined LLM outputs; (2) Diverse oracles beyond crashes are needed for LLM inference engine testing; (3) We reveal actionable diagnostic patterns for all identified bug symptoms.
Our research provides an empirical foundation for analyzing and testing LLM inference engines, with practical implications for different stakeholders and general guidelines for LLM inference engine development.